\documentclass[twocolumn]{aastex62} 
\usepackage{graphicx}
\usepackage{txfonts}
\usepackage{threeparttable}
\received{xxx}
\revised{xxx}
\accepted{}
\submitjournal{ApJ}
\shorttitle{UV/optical thermal reverberation in AGN}
\shortauthors{Kammoun et al.}

\def\rg{$R_{\rm g}$}

\def\cm3{\hbox{cm$^{-3}$}}

\def\mdot{\hbox{$\dot{m}$}}

\def\Lx{\hbox{$L_{\rm X,Edd}$}}
\def\tcen{\hbox{$\tau_{\rm cen}$}}
\def\Rfrac{\hbox{${\cal{R}}_{\rm rev}$}}

\begin{document} 

\title{UV/optical disk thermal reverberation in AGN: an in-depth study with an analytic prescription for the time-lag spectra}

\correspondingauthor{E. S. Kammoun}
\email{ekammoun@umich.edu}

\author[0000-0002-0273-218X]{E. S. Kammoun}
\affiliation{Department of Astronomy, University of Michigan, 1085 South University Avenue, Ann Arbor, MI 48109-1107, USA}

\author{M. Dov\v{c}iak}
\affiliation{Astronomical Institute of the Czech Academy of Sciences, Bo{\v c}n{\'i} II 1401, CZ-14100 Prague, Czech Republic}

\author{I. E. Papadakis}
\affiliation{Department of Physics and Institute of Theoretical and Computational Physics, University of Crete, 71003 Heraklion, Greece}
\affiliation{Institute of Astrophysics, FORTH, GR-71110 Heraklion, Greece}

\author{M. D. Caballero-Garc\'{i}a}
\affiliation{Astronomical Institute of the Czech Academy of Sciences, Bo{\v c}n{\'i} II 1401, CZ-14100 Prague, Czech Republic}

\author{V. Karas}
\affiliation{Astronomical Institute of the Czech Academy of Sciences, Bo{\v c}n{\'i} II 1401, CZ-14100 Prague, Czech Republic}
\begin{abstract}

Several active galactic nuclei show correlated variations in the ultraviolet/optical range, with time delays increasing at longer wavelengths. Thermal reprocessing of the X-rays illuminating the accretion disk has been proposed as a viable explanation. In this scenario, the variable X-ray flux irradiating the accretion disk is partially reflected in X-rays, and partially absorbed, thermalized and re-emitted with some delay by the accretion disk at longer wavelengths. We investigate this scenario assuming an X-ray point-like source illuminating a standard Novikov-Thorne accretion disk, around a rotating black hole. We consider all special and general relativistic effects to determine the incident X-ray flux on the disk and in propagating light from the source to the disk and to the observer. We also compute the disk reflection flux taking into consideration the disk ionization. We investigate the dependence of the disk response function and time lags on various physical parameters, such as black hole mass and spin, X-ray corona height, luminosity, and photon index, accretion rate, inclination, and inner/outer disk radii. We found it is important to consider relativistic effects and the disk ionization in estimating the disk response. We also found a strong non-linearity between the X-ray luminosity and the disk response. We present an analytic function for the time-lags dependence on wavelength, which can be used to fit observed time-lag spectra. We also estimate the fraction of the reverberation signal with respect to the total flux and we suggest possible explanation for the lack of X-ray-ultraviolet/optical correlated variations in a few sources.

\end{abstract}


\section{Introduction}
\label{sec:intro}

According to the current paradigm,  active galactic nuclei (AGN)  are thought to be powered by accretion of matter, in the form of a geometrically thin and optically thick disk, onto a central supermassive black hole (BH) of mass $M_{\rm BH} \sim 10^{6-9}~ \rm M_\odot $. The viscous accretion disk \citep{Pringle72, Shakura73, Novikov73} emits the bulk of its light in the ultraviolet(UV)/optical range, and is responsible for most of the observed bolometric luminosity. AGN are known to be strong X-ray emitters as well. The X-rays are thought to be produced in the close vicinity of the BH, by hot electrons ($\sim 10^9 ~\rm K$), located in a region which is known as the `X-ray corona'. Thermal UV photons arising from the accretion disk will be Compton up-scattered off the energetic electrons in the X-ray corona, giving rise to the `primary emission' \citep[e.g.,][]{Lightman88, Haardt93}. The primary emission spectrum is well described by a power law with a high energy cutoff. 

Part of the primary emission will be detected by the observer, and the other part will shine on the accretion disk. In the latter case, a fraction of the light will be reprocessed and emitted by the disk in the X-rays \citep[e.g.,][]{George1991, Matt1993}. The other fraction will be absorbed by the disk, increasing its temperature, and re-emitted in the form of thermal UV/optical radiation. This will increase the UV/optical flux of the disk. In the event of a variable X-ray flux, the additional thermalized UV/optical flux will also be variable, with a time lag (with respect to X-rays) increasing with wavelength. This mechanism is known as the disk thermal reverberation \citep[e.g.,][]{Cackett07}.

Various high-cadence monitoring campaigns, across X-rays, UV, and optical, have been performed during the last years, using space and ground-based telescopes \citep[e.g.,][]{Mchardy14, Shappee14, Mchardy18, Cackett18, Edelson19, Cackett20, Santisteban20}. In general, the UV/optical variations are well correlated, with the optical variations being delayed with respect to the UV. This is in agreement with the hypothesis of disk thermal reverberation.  

Recently we studied the disk thermal reverberation in the case of NGC~5548 \citep[herafter KPD19]{paper1}. We computed the disk response to the X-rays, and the corresponding time lags, assuming a standard Novikov-Thorne accretion disk \citep[NT;][]{Novikov73} illuminated by a point-like X-ray source, located above the BH (known as the `lamp-post' geometry), as a simplified representation of an on-axis compact corona. The model took into account relativistic effects and we computed the disk reflection, accounting for its ionization profile. We investigated the effects of the accretion rate and the height of the X-ray corona on the time lags between X-rays and UV/optical. We were able to fit the `lag vs. wavelength' plot (this is known as the `time-lags spectrum') that \cite{Fausnaugh16} previously estimated in NGC~5548, and we showed that the time-lags spectrum is in agreement with a NT disk, accreting at a small rate, as long as the X-ray source is at a distance larger than 40 gravitational radii ($R_{\rm g} = GM_{\rm BH}/c^2$) above the BH. 

In this work, we present the results from an extended study of the disk thermal reverberation in the case of the lamp-post geometry. We compute the general relativistic (GR) effects on the intrinsic X-ray spectrum of the corona, on the incident X-ray spectrum on the disk and the resulting X-ray reflection spectrum in more detail than KPD19. We show that there are significant differences in the disk response functions when we consider all the GR and disk ionization effects and when we use the approximations that have been usually assumed in the past. We compute and study the disk response for a wide range of values of model parameters -- BH mass, accretion rate, corona height, the energy spectral photon index, inclination angle, incident X-ray luminosity, and the disk inner and outer radius. We compute time lags as a function of wavelength for the full parameter space, and we determine an analytic function for the time-lags between X-rays and the UV/optical light curves that can be used to fit the observed time-lags at least up to 5000~\AA. We also compute the ratio of the thermally reverberating UV/optical disk flux over the underlying NT disk flux in various energy bands (the `reverberation fraction' hereafter). This ratio can be used to explain the non-detection of the variable, UV/optical reverberating component in AGN which are highly variable in X-rays. The paper is organized as follows. 

In Section~\ref{sec:model} we present the model setup, and we compare the model disk responses with the  responses when using the approximations that have been frequently adopted in the past. In Section~\ref{sec:response} we present the disk responses for a wide range of model parameter values. In Section~\ref{sec:application} we discuss possible implications of our results, and in particular, we compute time-lag spectra and we derive the analytic expression for the time-lags as a function of wavelength. We also discuss the reverberation ratio for the parameter space that we considered. We conclude with a short summary of our results.

\section{Model setup}
\label{sec:model}

Similar to KPD19, we consider a Keplerian, geometrically-thin and optically-thick accretion disk, around a BH of mass $M_{\rm BH} $ and accretion rate \mdot\ in the Kerr metric\footnote{The dimensionless spin parameter, defined as $a^\ast = Jc/GM^2$, where $J$ is the angular momentum of the BH, is smaller or equal to 1 \citep[see e.g.,][]{Misner73}.}. The disk temperature profile follows the NT prescription, with a color temperature correction factor of 2.4 \citep{Ross92}. As a model for the corona, we assume the lamp-post geometry: the X-ray source is point-like and is located at a height $h$ above the BH, on its rotational axis. 

In order to study the disk thermal reverberation when illuminated by the variable X-rays, we need to determine the disk response to an X-ray flash. We assume that the X-ray corona emits isotropically (in its rest frame) a power-law spectrum of the form $f_{\rm X}(E,t) = N(t)E^{-\Gamma}\exp(-E/E_{\rm cut})$, where $\Gamma$ is the spectral photon index (assumed to be constant). The flash lasts for $\Delta t=10~T_{\rm g}$ (where $T_{\rm g}$ is the light travel time of the gravitational radius, $R_{\rm g}$), and it has a top-hat shape (i.e., $N(t)$ has a constant, non-zero value for $0 \leq t \leq 10~T_{\rm g}$, and then $N(t)=0$ for $t > 10~T_{\rm g}$). Since we use numerical computations, we cannot assume an ideal $\delta$-function but we approximate the X-ray flash with the top-hat function instead with a carefully chosen width. This width is small enough to avoid inaccuracies in our computations of the responses (especially for the response duration and average response time) and at the same time it is large enough to achieve reasonable computing times to get response functions that are smooth enough (i.e., without numerically caused oscillations). The flash duration of $10~T_{\rm g}$ amounts to $5.7 \times 10^{-3}$~day for a black hole mass of $10^7~M_\odot$ thus being much smaller than the average response duration, which is usually measured in the order of days (i.e., smaller by more than 2 orders of magnitude). It is even smaller than the start time of the response (with the exception of low heights below $\sim 10\,R_{\rm g}$ and high inclinations above $\sim 60^\circ$). For more details we refer the reader to Figures~\ref{fig:mass}, \ref{fig:height} and \ref{fig:theta}.

Let us consider the total flux received by the disk, at radius $R$ from the center, at time $t$ after the start of the flash, $F_{\rm inc}(R,t)$. Part of this flux will be reflected and re-emitted in X-rays (this is the `disk reflection component') and part of it will be absorbed, as follows:

\begin{equation}
    F_{\rm abs} (R, t) = F_{\rm inc}(R, t) - F_{\rm ref}(R, t),
    \label{eq:fabs}
\end{equation}

\noindent where  $F_{\rm ref}(R, t)$ is the (total) reflected flux. The absorbed X-rays will thermalize in the disk, and will act as an extra source of heating. Consequently, the local disk temperature will increase. We can use the sum of $F_{\rm abs}(R,t)$ and the original, total NT flux emitted by the disk at radius $R$, $F_{\rm NT}(R)$ (which we assume is constant), to estimate the new disk temperature as follows:

\begin{equation}
    T_{\rm new}(R, t) = \left[\frac{F_{\rm abs}(R, t) + F_{\rm NT}(R)}{\sigma}\right]^{1/4},
\end{equation}

\noindent where $\sigma$ is the Stefan-Boltzmann constant. 

The disk response function in a waveband\footnote{ A waveband is determined by its centroid wavelength, $\lambda_{\rm c}$, and its width $\Delta\lambda=\lambda_{\rm max}-\lambda_{\rm min}$.}, $\Psi (\lambda_{\rm c}, t_{\rm obs})$, is defined in such a way so that it is equal to the flux that the disk emits due to X-ray heating at time $t_{\rm obs}$ (as measured by a distant observer). This flux varies with time because the X-ray flash first illuminates the inner disk and then propagates to the outer parts. To determine the disk thermal response, (i) we identify all the disk elements that brighten up at $t_{\rm obs}$, thus correspondent to the observed flash reflection image on the disk $[R,\phi]_{t_{\rm obs}}$, (ii) we compute the sum of their thermal flux, $F^{\rm flash}_{\rm tot}(\lambda_{\rm c}, t_{\rm obs})$ \footnote{$F^{\rm flash}_{\rm tot}$ is the sum of the blackbody flux of disk elements, each with its own temperature $T_{\rm new}$.}, and (iii) we subtract the sum of their intrinsic disk flux, $F^{\rm flash}_{\rm NT}(\lambda_{\rm c}, t_{\rm obs})$ \footnote{The NT disk flux is constant with time, but $F^{\rm flash}_{\rm NT}$ changes with time since it corresponds to the NT thermal flux produced by the evolving reflection image on the disk $[R,\phi]_{t_{\rm obs}}$}, so that 

\begin{equation} \label{eq:psi}
    \Psi(\lambda_{\rm c}, t_{\rm obs}) = \frac{F^{\rm flash}_{\rm tot}(\lambda_{\rm c}, t_{\rm obs}) - F^{\rm flash}_{\rm NT}( \lambda_{\rm c}, t_{\rm obs})}{L^{\rm flash}_{\rm Xobs, Edd}~\Delta t},
\end{equation}
\noindent where $\Delta t$ is the duration and $L^{\rm flash}_{\rm Xobs,Edd}$ is the observed, 2--10~keV luminosity of the X-ray flash (in Eddington units). The disk response is normalized to the X-ray luminosity so that the observed flux (in the UV/optical bands), when the disk is constantly being illuminated by variable X-rays, will be given by,

\begin{equation}\label{eq:Fobs}
    F_{\rm obs}(\lambda_{\rm c}, t) = F_{\rm NT}(\lambda_{\rm c}) + \int_{0}^{+\infty} L_{\rm Xobs,Edd}(t-t')\Psi(\lambda_{\rm c}, t'){\rm d}t'. 
\end{equation}
\noindent Here, $F_{\rm NT}(\lambda_{\rm c})$ is the NT emission  from the whole disk in the waveband with centroid wavelength $\lambda_{\rm c}$,  $L_{\rm Xobs,Edd}(t)$ is the observed, 2--10~keV luminosity of the corona, and the convolution in the right hand side of the equation above gives the variable, thermally reprocessed disk flux. 

In Section \ref{sec:response} we present model disk responses for various model parameter values, using Equation~(\ref{eq:psi}). But first, we discuss modifications to the model with respect to KPD19. 

\subsection{The $low$-energy cut-off in the X-ray spectrum}
\label{sec:Xrayemission}

\begin{figure}
\centering
\includegraphics[width=0.95\linewidth]{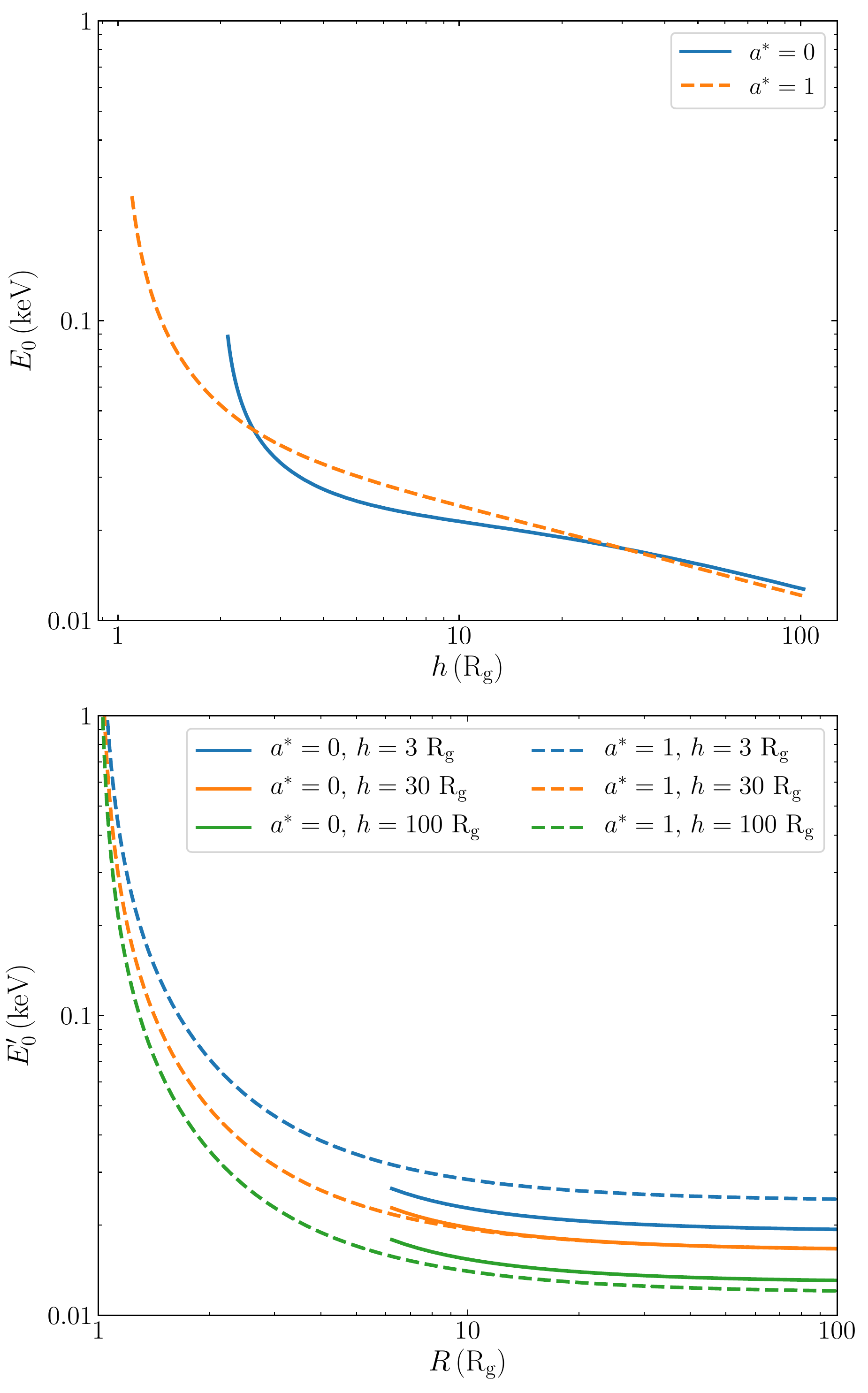}

\caption{Upper panel: Dependence of the intrinsic rollover energy, $E_0$, on the X-ray corona height. Bottom panel: Plot of the rollover energy, $E'_0$, as seen by the disk, as a function of radius, for several heights. The solid and dashed lines correspond to  $a^{\ast}=0$ and 1, respectively. The energies are estimated for a $10^6$ $M_{\odot}$ black hole, and an accretion rate of 0.01 $\dot{m}_{\rm Edd}$.}
\label{fig:E0}
\end{figure}

The X-ray spectrum depends on four parameters: i) normalization (which is defined by $L^{\rm flash}_{\rm Xobs,Edd}$), ii) the power-law index, $\rm \Gamma$; iii) the exponential cut-off at high energy ($E_{\rm cut}$), and iv) the $low$ energy rollover, $E_0$, which is determined by the typical energy of the seed photons, as seen by the corona. All the four parameters should be known to determine the total luminosity of the corona, and thus, the incident X-ray flux, $F_{\rm inc}(R,t)$, at each disk radius. $L^{\rm flash}_{\rm Xobs,Edd}$, $\Gamma$ and $E_{\rm cut}$ are `independent' parameters, in the sense that their values depend on the assumed corona internal properties (e.g., corona size, optical depth, electron temperature). However, this is not the case for $E_0$, since it depends on the accretion rate and on the corona height, $h$ (for a given BH mass). KPD19 assumed $E_0=0.1$~keV. However, in this paper we compute its exact value  for the spectrum emitted by the corona and also for the spectrum seen by each disk element.

 We follow the same approach as in \cite{Dovciak2016}, and we compute the disk thermal spectrum at the location of the corona, integrated over all directions. This is a multi-color blackbody (BB) spectrum (with color factor correction included), yet it can be well approximated with a single BB except for the fading tail towards higher energies. This is due to the fact that the disk inner parts, with the highest temperature, contribute the most to the disk emission arriving at the corona, while the outer, colder disk regions illuminate the corona with much smaller solid angle. We use the temperature of this single BB approximation, $T_{\rm BB}$, and we compute the typical seed photon energy as $E_0=kT_{\rm BB}$. Therefore, $E_0$ depends  on $M_{\rm BH}$ and $\dot{m}$, as these parameters determine the disk effective temperature in the first place, hence $T_{\rm BB}$ as well. In the case of a NT disk which extends to the innermost stable circular orbit ($R_{\rm ISCO}$), $E_0$ scales as, 

\begin{equation} \label{eq:e0}
 E_0 (M_{\rm BH},\dot{m}_{\rm Edd},h, a^\ast) \propto \left[\frac{\dot{m}/\dot{m}_{\rm Edd}}{0.01} \frac{1}{M_6} \right]^{1/4},  
\end{equation}

\noindent where $M_6$ is the BH mass in units of $10^6~ \rm M_\odot$ and $\dot{m}_{\rm Edd}$ is the Eddington accretion rate. The normalization in this equation depends on the height of the corona and the BH spin parameter. The dependence on height comes mainly due to  gravitational redshift, since the energy of the emitted disk photons decreases as they escape the BH. It also depends on the transverse Doppler shift since the corona is at rest while matter in the innermost regions of the disk is orbiting very fast (with velocities that can be close to the speed of light). In general, $E_0$ should decrease with increasing $h$. The dependence on the spin is due to the different energy of the seed photons emitted by the disk since the disk inner edge and the physical value of \mdot\ (in $\rm M_\odot/yr$) changes with the BH spin\footnote{We remind the reader that the radiative efficiency $\eta$ is smaller for low spins, hence $\dot{m} = L/\eta c^2$ in physical units is higher for low spins.}.

The top panel of Figure~\ref{fig:E0} shows the plot of $E_0$ at the corona as a function of $h$, for $\dot{m}/\dot{m}_{\rm Edd} = 0.01$, $M_{\rm BH} = 10^6~\rm M_\odot$, and $a^\ast = 0,$ and 1 (solid and dashed curves, respectively). The $E_0$ values plotted in the figure set the normalization in Equation~(\ref{eq:e0}). The plot shows that, for this combination of BH mass and accretion rate, $E_0$ is smaller than 0.1~keV (the value  KPD19 assumed), at all heights (except for $h\lesssim 1.5\,R_{\rm g}$, when $a^\ast=1$). 

The rollover energy seen by the disk, $E'_0 (R)$, will be further modified due to photons energy shift, as they travel from the corona to the disk elements. The bottom panel of Figure~\ref{fig:E0} shows $E'_0$ as a function of the disk radius for several corona heights. One can see that its value is below 0.1~keV for the assumed BH mass and accretion rate, except very close to the black hole ($R\lesssim 2\,R_{\rm g}$) in the case of a high BH spin, where it can reach $\sim 0.1$~keV (but close to the horizon even up to $\sim 1$~keV).

\subsection{The $high$-energy cut-off in the X-ray spectrum}

KPD19 used the {\tt xillverD} tables \citep{Garcia16}, which allow the use of different disk densities, but the high-energy cutoff ($E_{\rm cut}$) is fixed at 300~keV. In reality, the intrinsic high-energy cut-off can be smaller or larger than 300~keV. Even if it is 300~keV, $E'_{\rm cut}(R,t)$ in the rest frame of a disk ring can be shifted to very high energies for rings close to the event horizon in the case of large heights. It can be also shifted to much lower energies for rings which are far from the BH in case of very low heights. Hence, if $E_{\rm cut}$ is fixed, the incident flux can be over or underestimated, especially for hard energy spectra (i.e., $\rm \Gamma<2$). Therefore, an alternative possibility is to consider the {\tt xillver-a-Ec5} \citep{Garcia2013,Garcia2014} tables which are computed for a single disk density of $10^{15}~\rm cm^{-3}$, and various $E_{\rm cut}$ values. 

\begin{figure}
\centering
\includegraphics[width=0.95\linewidth]{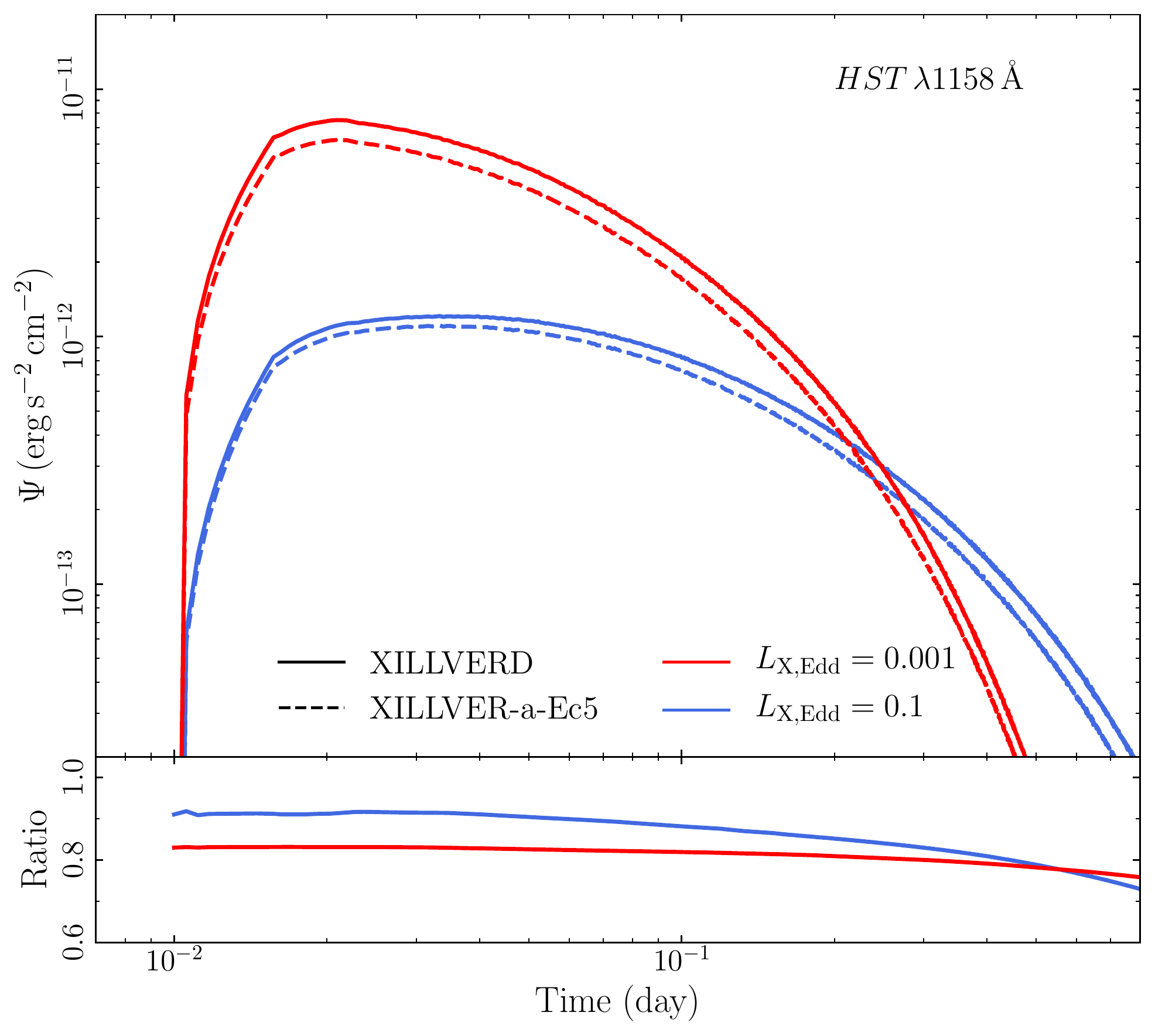}
\caption{Response functions in the {\it HST} $\lambda 1158$\AA\, band, using the {\tt XILLVER-D} and {\tt XILLVER-a-Ec5} tables with $n_{\rm H} = 10^{15}~\rm cm^{-3}$(solid and dashed lines respectively), in the case when $a^\ast = 1$. We assumed $L_{\rm Xobs, Edd} = 0.001$ and 0.1 (red and blue lines, respectively). The lower panel shows the ratio of the response functions.}
\label{fig:nh}
\end{figure}

To investigate the significance of  $E'_{\rm cut}(R,t)$ we used the fiducial parameter values listed in Table \ref{table:param}, and we computed the disk response function in the first waveband listed in Table \ref{table:bands}, for $L^{\rm flash}_{\rm Xobs,Edd}=0.001$ and $0.1$. The results are plotted in Figure\,\ref{fig:nh}. The solid and dashed lines show the response when we use the {\tt xillverD} and the {\tt xillver-a-Ec5} tables, respectively ($n=10^{15}$ cm$^{-3}$, in both cases; the amplitude of the response function is not the same for the two X-ray flash luminosities for reasons we explain in Section \ref{sec:Lx}).

The amplitude of the response is larger when we use the {\tt xillverD} tables, at all times, both for the high and the low X-ray illuminating flux. This is because the fiducial (observed) cut-off energy is 150~keV, which implies an intrinsic cut-off of $\sim 180$~keV for a corona located at $h=10 ~R_{\rm g}$ \citep[see][]{Tamborra18}. This is much smaller than the default intrinsic cut-off of 300~keV in the case of the {\tt xillverD} tables, so we consistently overestimate the incident and the absorbed flux when using these tables. 

The difference between the two responses is of the order of 10-20\%, but it will increase for harder spectra (i.e., $\Gamma<2$) or cut-off energies smaller than 150~keV. The difference in the amplitude of the response function will reverse when the intrinsic cut-off energy is larger than 300~keV. In this case, using the {\tt xillverD} tables we will underestimate the intrinsic and absorbed flux, and the response amplitude will be smaller than what it should be. For these reasons, we decided to compute the disk responses with the {\tt xillver-a-Ec5} tables (and fix the disk density at 10$^{15}$ cm$^{-3}$).

\subsection{The disk reflection spectrum}
\label{sec:reflection}

In addition to $F_{\rm inc}(R,t)$, we also need to know the disk reflection flux, $F_{\rm ref}(R,t)$, to accurately compute the absorbed X-ray flux. To compute $F_{\rm ref}(R,t)$ we determine the disk ionization (which depends on $F_{\rm inc}(R,t)$ and the disk density, $n$), and then we use the appropriate {\tt xillver} tables. 

 The {\tt xillver} tables provide the X-ray reflection spectrum at energies above 0.1~keV. If the low-energy cut-off in the rest frame of each disk ring, $E_0'(R,t)$, is smaller than 0.1~keV, we assume that the incident flux below 0.1~keV is fully absorbed, and we compute $F_{\rm ref}(R,t)$ by integrating the {\tt xillver} tables above 0.1~keV (as in KPD19). In some rings though, $E_0'(R,t)$ is larger than 0.1~keV. This can happen close to the black hole (depending on $M_{\rm BH},~ \dot{m}$ and $h$, see Equation~(\ref{eq:e0}) and Figure~\ref{fig:E0}). We considered two options to address this issue: 1) use the {\tt xillver} tables with their original cutoff at 0.1~keV, and 2) integrate the {\tt xillver} tables above $E_0'(R,t)$. In the first case we risk overestimating the reflection flux, while in the second approach we could underestimate it. We tested several cases with both approaches and we found that the differences in the resulting response functions were negligible. In the rest of the analysis, we estimate the response functions by integrating the {\tt xillver} tables above $E_0'(R,t)$, whenever it is larger than 0.1~keV.
\begin{figure*}
\centering
\includegraphics[width=0.9\linewidth]{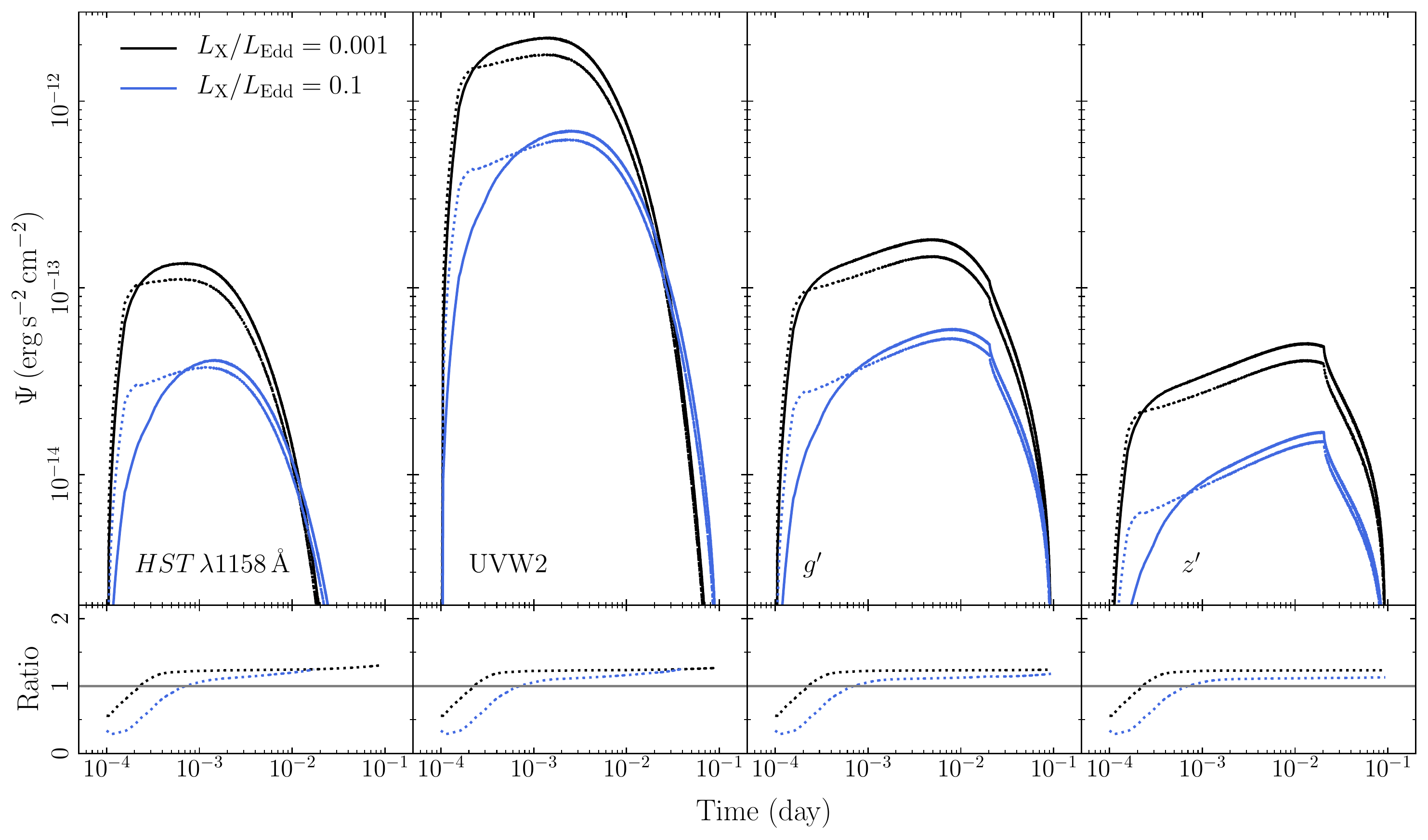}
\caption{Top panels: Response functions for $L_{\rm Xobs,Edd} = 0.001$ and 0.1 (black and blue, respectively) when the disk reflection spectrum is taken into account including the disk ionization (solid lines) and when assuming an albedo of 0.3 (dotted lines). Bottom panels: the ratio of the response functions for each luminosity (See Section~\ref{sec:reflection-albedo} for details). }
\label{fig:albedo}
\end{figure*}
\subsection{Comparison with past studies}
Before presenting our results, we investigate whether our approach in computing $F_{\rm inc}(R,t)$ and $F_{\rm ref}(R,t)$ results in response functions which are significantly different when compared to the response functions computed with various approximations that have been commonly assumed in the past.

\subsubsection{Constant albedo vs self-consistent reflection from ionized disk}
\label{sec:reflection-albedo}

In most cases, the disk reflection flux in the past was calculated by assuming a constant disk albedo. 
To test the importance of computing the disk reflection flux taking into account the disk ionization, as opposed to assuming a constant disk albedo, we considered model parameters in such a way so that $E_0$, as seen by the corona and disk, is as large as possible. In this case, we can compute accurately the disk reflection flux, and hence the amount of thermalization, as the incident flux below $0.1\,$keV (down to which the reflection tables exist) will be minimal. 

Thus, we chose a rather extreme case with $M_{\rm BH} = 10^5 ~\rm M_\odot$ and $\dot{m}/\dot{m}_{\rm Edd} = 0.1 $, when the disk temperature, and consequently the seed photon energy, will be high. To see the effects for different disk ionization states we further set $L_{\rm Xobs, Edd} = 0.001$ and $0.1$. For the rest of the parameters we choose $\Gamma = 2$ for the primary power-law photon index, $h = 10~R_{\rm g}$ for the corona position, $a^{\ast}=1$ for the BH spin and $\theta=40^\circ$ for the system inclination. We consider the case when the reflection flux is computed in a self-consistent way, and the case when we fix the thermalized flux to be 70\% of the X-ray incident flux (equivalent to considering a constant disk albedo of 0.3). All GR effects are taken into consideration in computing correctly the incident X-ray flux in both cases (i.e., $E'_0(R,T)$, $E'_{\rm cut}(R,t)$, and hence $F_{\rm inc}(R,t)$, are computed in the same way in both cases). The results are shown in Figure~\ref{fig:albedo} for four different wavebands. 

This plot shows that computing the reflection spectrum by assuming an albedo of 0.3 overestimates $\Psi$ at early times when $\Psi$ rises, and underestimates it later. This effect is more important when $L_{\rm Xobs, Edd}$ is high, as the gradient in the ionization profile of the disk is more important compared to lower X-ray luminosity. It is worth noting that a smaller albedo (i.e., larger thermalization fraction) will increase the difference seen at early time as it will predict a larger $\Psi$. The difference in the response function in the two cases will be important in the prediction of the reverberation signal in the various UV/optical wavebands, as the case of a constant albedo will overestimate the predicted flux, especially in the case when the incident X-ray flux is large. 

\subsubsection{The X-ray incident spectrum vs fixed X-ray incident flux}
\label{sec:fixedornotLx}

\begin{figure*}
\centering
\includegraphics[width=0.9\linewidth]{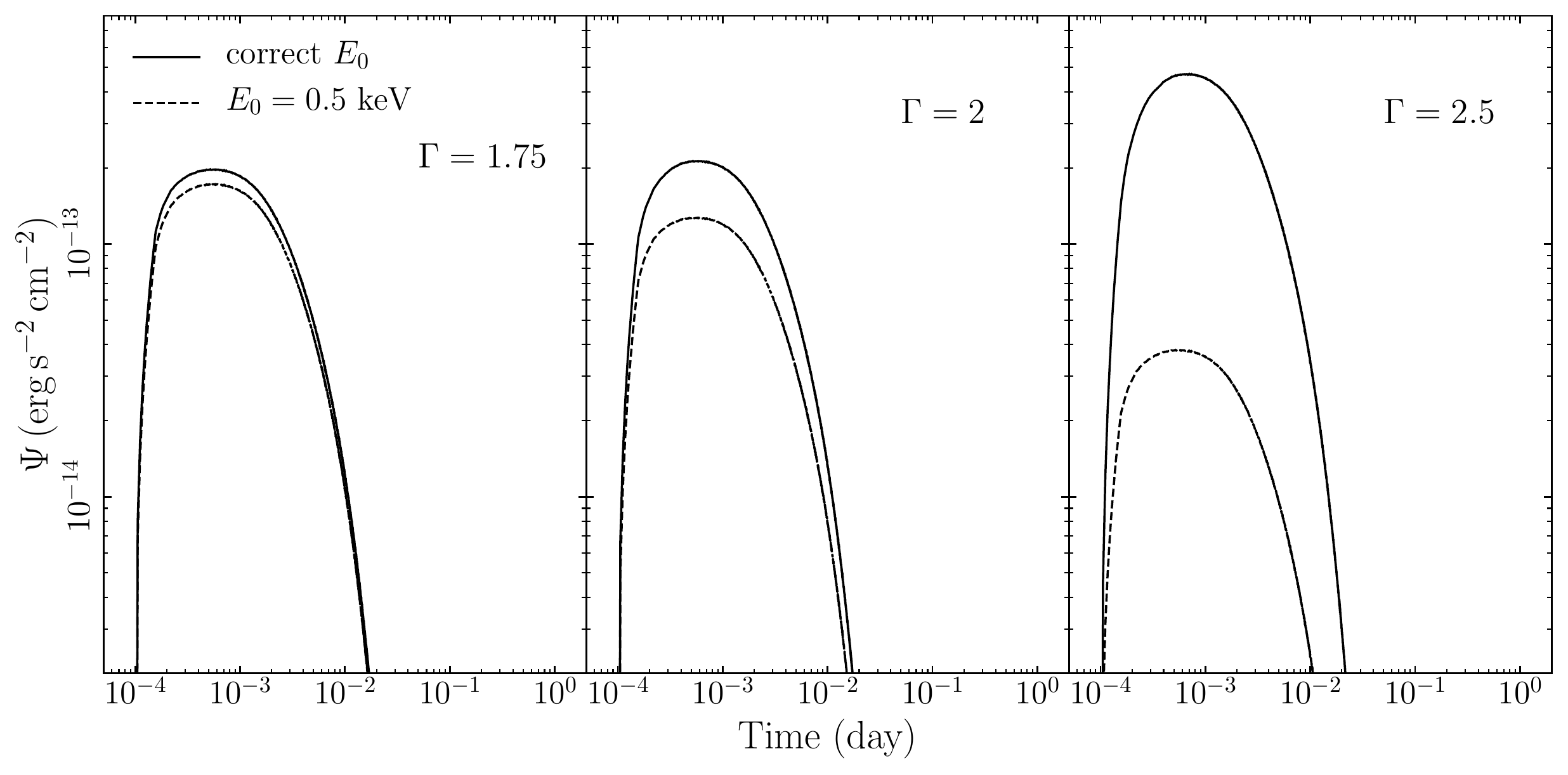}
\caption{The response functions, in the \textit{HST}$\lambda 1158\,$\AA\/ band, estimated by fixing $E_0$ at 0.5~keV (dashed lines) and by estimating $E_0$ correctly. The power-law normalization is set in a way that the total X-ray luminosity is the same in both cases. We show the response functions for different values of $\Gamma$ (increasing from left to right).}
\label{fig:diffSpec}
\end{figure*}

In most cases in the past, a fixed flux for the X-ray corona was assumed. However, as we mentioned in Section \ref{sec:Xrayemission}, the low and high energy cut-offs in the spectrum of the X-ray corona also play a role in determining the X-ray flux that will be absorbed by the disk.  In order to highlight the importance of the spectral shape of the primary emission in determining accurately the response functions, we performed the following test. 

We considered a BH with a mass of $10^5$M$_{\odot}$ and $L_{\rm Xobs,Edd}=0.001$, and the other model parameters set to the fiducial values listed in Table~\ref{table:param}. We considered two illuminating X-ray spectra, having the same photon index, and the same $E_{\rm cut}$. In the first case we set $E_0$ (in the intrinsic corona energy spectrum that shines the disk) to be the correctly estimated value (as we described in Section~\ref{sec:Xrayemission}). In the second case we fixed $E_0$ at 0.5~keV. The normalization of the two energy spectra is set to ensure the total X-ray luminosity (between $E_0$ and 1~MeV) is 0.001 (in Eddington units), in both cases.

The response functions for the \textit{HST}~$\lambda 1158\,$\AA\/ are shown in Figure~\ref{fig:diffSpec}. The correct $E_0$, for the model parameters we chose, is lower than 0.5 keV. As a result, the amplitude of the response function is smaller when $E_0$ is fixed at 0.5~keV, because less soft X-rays are absorbed and thermalized than in reality. This effect is more pronounced in the case of steep photon indices, where the amplitude of the response functions can differ by a factor of 10 or more. Consequently, a fixed X-ray luminosity, without considering all the parameters of the X-ray spectrum, may lead to the calculation of the wrong disk response.   

\subsubsection{The significance of GR effects}

Finally, we investigate the effects of accounting for general relativity (GR) when estimating the photon trajectory from the X-ray source to the disk, and also when computing $E_0$ and $E'_0(R,t)$. We consider the fiducial model parameters (see Table \ref{table:param}) and two cases for the corona height: $h=2.5$~\rg\ and 100~\rg. We computed the response functions following the same steps, but in the first case using full GR computations, while in the second case we take Newtonian approach with photons moving along straight lines and not shifting their energy. In the second case we still assume $E_0$ to be equal to the temperature of the disk spectrum approximation we described in Section~\ref{sec:Xrayemission}, while the intrinsic $E_{\rm cut}$ to be the same as the observed one. We also take into account the disk ionization effects in both cases.

\begin{figure*}
\centering
\includegraphics[width=0.95\linewidth]{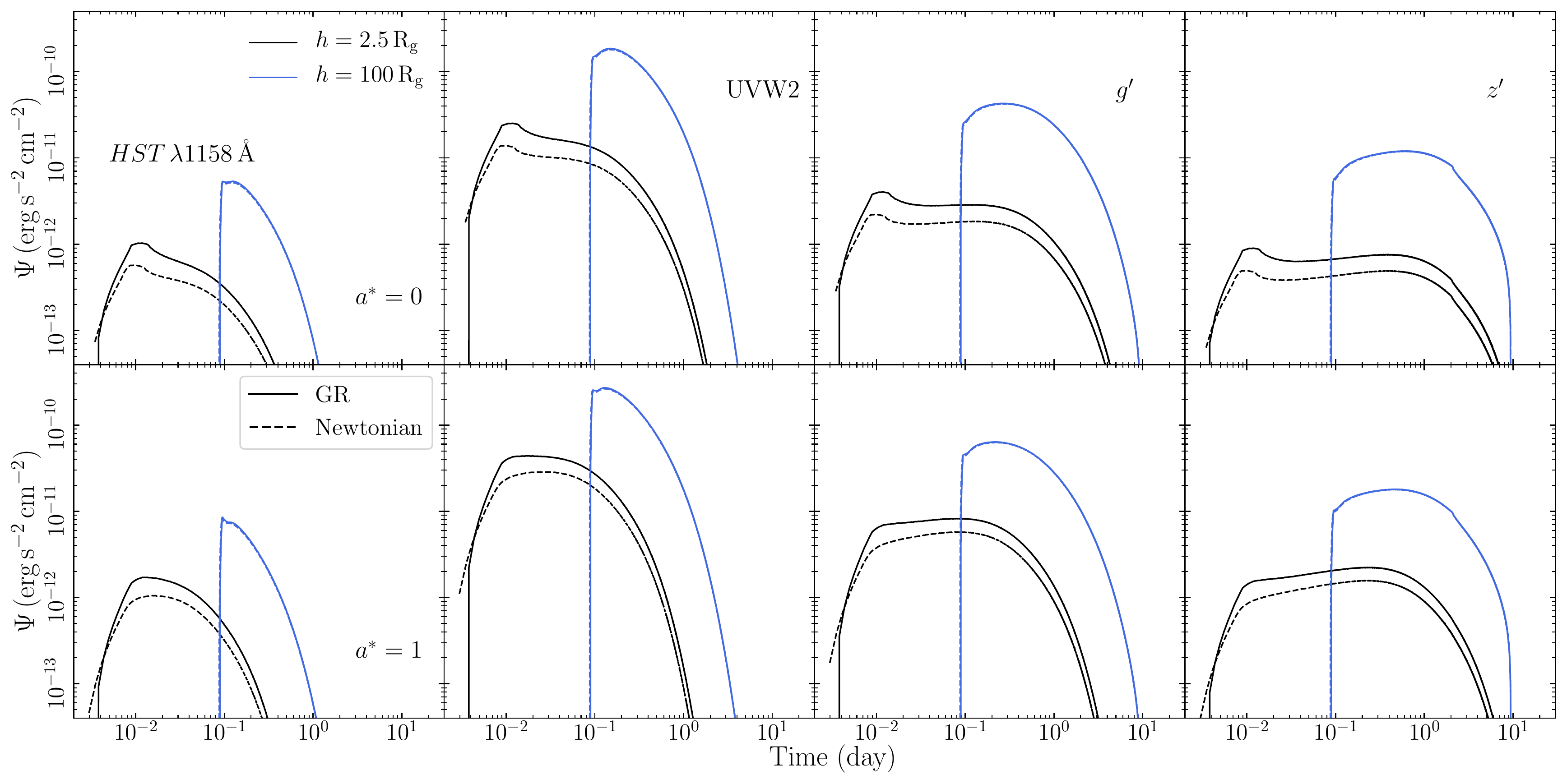}

\caption{Response functions assuming fiducial parameters and $h = 2.5~R_{\rm g},$ and 100~\rg\ (black and blue, respectively) assuming general relativity (solid lines) and the Newtonian approximation (dashed lines) for $a^\ast = 0, 1$ (top and bottom panels, respectively). }
\label{fig:GR}
\end{figure*}

The response functions are presented in Figure~\ref{fig:GR}. The figure shows that the difference between the two approximations is minimal for the large height. However, for $h=2.5$~\rg, the response functions are smaller in amplitude in the Newtonian approximation. This implies that the Newtonian approximation underestimates the reverberating signal in the UV/optical bands. In addition, the Newtonian transfer functions rise (slightly) earlier than the case when we consider GR.

\section{The disk response functions}
\label{sec:response}

Table \ref{table:param} lists all the physical parameters that influence the disk response to an illuminating X-ray source. Our aim is to investigate how these parameters affect the disk response. We used Equation~(\ref{eq:psi}) and we computed the disk response for a broad range of values for each parameter, which are listed in the right column of Table~\ref{table:param}. The numbers in blue indicate the fiducial parameter values: we keep the parameter values fixed to these numbers when we vary another parameter. For each combination of parameter values, we compute the response for a non-rotating and a maximally spinning BH ($a^{\ast}=0$ and 1, respectively). 

We computed the response function in 10 wavebands, as listed in Table~\ref{table:bands}. The first three are the bands where the far-UV continuum was measured in NGC~5548, using the HST data taken during the Space Telescope and Optical Reverberation Mapping Project \citep[STORM;][]{Derosa15, Edelson15, Fausnaugh16}.  The next two bands are the Swift/UVOT, $W1$ and $W2$ bands, and the remaining 5 bands are representative of the SDSS $u',\,g',\,r',\,i',\,z'$ filters. The exact central wavelength, $\lambda_{\rm c}$, and width of the bands are not important at the moment. What is important is that the $\lambda_{\rm c}$'s are roughly equidistant from the far-UV to the near-IR. Consequently, the respective responses are representative of the disk response over the full UV/optical spectral band. We plot the response functions for all parameter values in Appendix  (Figures~\ref{fig:mass}-\ref{fig:Lx}). For clarity reasons, we show the response in 4 bands only (HST $\lambda 1158$, UVOT/$W2$, $g'$ and $z'$), which are representative of the disk response in the far-UV, UV, optical and near-IR. 

In the following subsections we investigate the dependence of the main properties of the response function, its start time, width and amplitude on the basic parameters of the system.

\begin{table}
\centering
\caption{The parameter values we used to compute the response functions. Numbers in blue and bold fonts are the fiducial values.}
\begin{tabular}{lll}
\hline \hline
Parameter	&		&	Range	\\ \hline
BH spin	&	$a^\ast$	&	{\bf \color{blue}  0, 1}	\\
BH mass	&	$M_{\rm BH}~(\rm 10^8~M_\odot)$	&	0.01, 0.05, {\bf \color{blue} 0.1}, 0.5, 1	\\
Accretion rate	&	$\dot{m}/\dot{m}_{\rm Edd}$	&	0.005, 0.01, 0.02, {\bf \color{blue} 0.05}, \\  
& & 0.1, 0.2, 0.5	\\
Corona height	&	$h ~( R_{\rm g})$	&	2.5, 5, {\bf \color{blue} 10}, 20, 40, 60, \\
 & & 80, 100	\\
Photon index	&	$\Gamma$	&	1.5, 1.75, {\bf \color{blue} 2}, 2.25, 2.5	\\
Inclination angle$^1$	&	$\theta$ (deg.)	&	5, 20, {\bf \color{blue} 40}, 60, 80	\\
Inner radius	&	$R_{\rm in} (R_{\rm g})$	&	{\color{blue} $ \bf R_{\rm isco}$}, 20, 100	\\
Outer radius    &	$R_{\rm out} (R_{\rm g})$	&	5$\times 10^2$, $10^3$, 5$\times 10^3$, {\color{blue} $ \bf 10^4$}	\\
X-ray luminosity	&	$L^{\rm flash}_{\rm Xobs, Edd}$	&	0.001, 0.0025, 0.005, {\bf \color{blue} 0.01},\\
& & 0.025, 0.05, 0.1, 0.25, 0.5	\\ 

\hline \hline 
\end{tabular}
$^1$ The angle between the normal to the disk and the line of sight.
\label{table:param}
\end{table}

\begin{table}
\centering
\caption{The central wavelengths and width for the different wavebands considered in this work.}
\begin{tabular}{lll}
\hline \hline
Band	&	$\lambda_{\rm c} $~(\AA)	&	$\Delta \lambda$~(\AA)	\\ \hline
$HST1158$	&	1158	&	5	\\
$HST1367$	&	1367	&	5	\\
$HST1746$	&	1746	&	5	\\
UVW2	&	1950	&	600	\\
UVW1	&	2600	&	700	\\
$u'$	&	3580	&	339	\\
$g'$	&	4754	&	1387	\\
$r'$	&	6204	&	1240	\\
$i'$	&	7698	&	1303	\\
$z'$	&	9665	&	2558	\\ \hline \hline
\end{tabular}
\label{table:bands}
\end{table}

\subsection{The response start time} 

The response start time, $t_{\rm start}$, is the same in all bands because we first detect the inner parts of the disk emitting BB radiation with a temperature that corresponds to a $\lambda_{\rm max} < 1158~$\AA. This temperature is high enough to contribute even in the shortest wavelengths we consider. The flux in the optical bands is significant even at early times, despite the fact that the disk area we observe at the beginning is small. This is due to the fact that the temperature of the inner disk is quite high (and the flux is proportional to $T^{4}$).

The disk responds to the X-ray flash at the same time, at all bands, when the geometry is fixed, i.e., for a given BH mass, corona height and inclination. The response start time, measured in days (and not in $T_{\rm g}=R_{\rm g}/c$), increases linearly with increasing $M_{\rm BH}$ (see also Figure \ref{fig:mass}). This is because all geometries studied and thus all distances and corresponding light travel times are scaled with $R_{\rm g}$ and $T_{\rm g}$, respectively, both of which scale with the black hole mass. The response start time also increases with the corona height, $h$, because the light travel between the corona and the disk increases accordingly. The relation between $t_{\rm start}$ and $h$ is almost linear, but not exactly (Figure \ref{fig:height}). This is due to the fact that photons do not travel along straight lines, especially when $h$ is small, and so the light-travel time does not depend linearly on $h$. 

 The observer sees the disk to respond to the X-ray flash much faster if the inclination is high (Figure \ref{fig:theta}). The observed $t_{\rm start}$ does not change significantly, as long as $\theta < 40^\circ$. But, at higher inclinations, $t_{\rm start}$ decreases fast as $\theta$ increases. Although GR contributes to this effect, it arises purely due to same geometrical reasons as in the flat space-time. For higher inclinations, the region that the observer sees to respond the earliest moves outward from the center. The difference between the length of the light path for photons coming to the observer directly from the source and those that are emitted by the disk in this region is much smaller for high inclinations.

It is interesting that the response start time does not depend on the BH spin. For all the parameter values we considered, $t_{\rm start}$ is identical in the $a^{\ast}=0$ and $a^{\ast}=1$ cases. This implies that we first detect disk elements that are located at a distance larger than $6~R_{\rm g}$, even for the lowest corona height we considered ($h=2.5 ~R_{\rm g}$) for the fiducial inclination of $40^\circ$.

\subsection{The width of the response} 
\label{sec:Psiwidth}

The width of the response increases at longer wavelengths. This effect was already noticed by KPD19. The response function in a particular band, $\lambda_{\rm c}$, starts decreasing when the temperature of the reverberating disk elements is such that the peak wavelength of their BB emission is longer than $\lambda_{\rm c}$. When this happens, it is the Wien part of the BB spectrum that contributes to the observed flux in this waveband, and it is therefore significantly decreased. As time passes, we detect emission from the outer, thus colder, parts of the disk. They can still contribute significantly in the optical bands, but not in the far-UV bands. Consequently, the width of the response is larger at longer wavelengths. In general, when the overall temperature of the disk is low, the disk response at a given waveband will be shifted closer to the black hole and thus the response will start decreasing at earlier times, becoming narrower. Therefore, any change in the system parameters that decreases the disk temperature causes the response width to be shorter. 

The discussion above explains the dependence of the response's width on the BH mass and the accretion rate. The width increases with increasing BH mass, but the rate of the increase in the response width with BH mass is smaller than the respective linear increase in $t_{\rm start}$. This is because, as $M_{\rm BH}$ increases, the disk temperature decreases. The same reason also explains the narrowing the response function as the accretion rate decreases, and the overall disk temperature drops (Figure~\ref{fig:mdot}).

As discussed at the beginning of this section, the response at each waveband depends on the disk temperature and thus it is confined to a disk region up to some radii where the response reaches the Wien part of the spectrum. If we increase the height, the reverberating image of a flash moves through this region faster. Therefore, intuitively one would expect that the width of the response should decrease by increasing the height. However, the opposite is true, and we observe the response width to increase with increasing the corona height (see Figure~\ref{fig:height}). The wavelengths we consider in this work are long enough so that regions with quite large radii still contribute to the response. For example, it is clear from Figure~\ref{fig:Rout} that, for $h=10~R_{\rm g}$, the disk elements just above $1000~R_{\rm g}$ still contribute to the responses even at the shortest studied wavelength of $1158$~\AA. For the higher heights even farther regions of the disk will contribute to the responses at short wavelengths. Since we explore heights below $100~R_{\rm g}$ (i.e. heights smaller than the maximum radius of the  reverberating region), the above described effect of shortening the response is quite small. 

There is, however, another effect in action that enhances the thermalized flux, and eventually leads to the increase in the response width with increasing height. The incident flux reaching the disk at large distances $R$ decreases as $1/(h^2+R^2)$ and is proportional to the cosine of the incident angle, which is defined as the angle between the normal to the disk and the photon trajectory. The response width depends on the disk emission at large distances, when $R\gg h$. Note, that for all the studied wavelengths, it is the contribution from regions $R>1000\,R_{\rm g}$ that is crucial for the width of the response, and the maximum height we considered is 100~$R_{\rm g}$, thus $R>10\,h$, at least. At these distances, the incident flux is independent of $h$ (as $1/(h^2+R^2)\sim 1/R^2$), and the cosine of the incident angle is equal to $h/\sqrt{h^2+R^2}\sim h/R$. Therefore, as $h$ increases, so does the cosine of the incident angle, and the incident flux as well. Due to the enhancement of the thermalized flux, disk regions further out will contribute more, thus effectively prolonging the response time (this effect increases towards longer wavelengths). Disk elements at small radii, where the cosine of incident angle decreases with increasing height, contribute only at the start time of the response function. Hence they do not affect the width of the response functions. 

The change in photon index, $\Gamma$, while keeping the X-ray luminosity in the $2-10$~keV band the same, changes the shape of primary power-law. When $\Gamma$ steepens, the flux below 1~keV increases. Since the low energy cut-off is usually below 0.1~keV, then the increase in $\Gamma$ increases the flux at low energies that contribute to the disk thermalization. As with the corona height, this effect leads to the response lasting longer (further out regions of the disk start to contribute to the thermal reverberation signal in all bands, even the shortest ones), as seen in Figure~ \ref{fig:gamma}. One can see that the largest change is when the power-law index is the steepest. The contribution of the primary radiation to thermalization below 0.1~keV changes very slowly for flat $\Gamma$.

The width of the response increases with larger inclination angle (see Figure~\ref{fig:theta}). This is due to the fact that it takes longer to detect the emission from the far side of the disk (due to geometrical reasons, a similar explanation but with an opposite effect as in the case of $t_{\rm start}$).

\subsection{The amplitude of the response}
\label{sec:Psiamplitude}
We recall that the response function gives the flux of the variable, thermally reverberated disk component (normalized to the illuminating X-ray flux) as a function of time. The integral of $\Psi(\lambda_{\rm c},t)$ over time is equal to the disk flux due to the X-ray heating (in each waveband). The response amplitude increases when the contribution of thermalization to the thermal flux at a given wavelength increases. This can be caused by several effects: (1) an increase of the total incident flux, (2) a decrease of the original NT thermal emission (i.,e. by decreasing the disk temperature) while keeping the total incident flux the same, or (3) a small ionization of the disk that then reflects less. We explore below how the various system parameters may contribute to these effects.

The BH mass can affect the response amplitude in various, and partially opposing, ways: i) The BH mass scales the source luminosity that is given in units of $L_{\rm Edd}$ and thus the normalization of the incident flux (by a factor proportional to $M_{\rm BH}$). (ii) A larger BH mass decreases the temperature of the disk and so increases the response amplitude. The lower disk temperature also implies a lower $E_0$ and the same X--ray luminosity in the $2-10$~keV band will lead to a larger total incident flux. Since $E_0$ is usually smaller than  0.1~keV, this additional flux will mainly contribute to the thermalized part, effectively increasing the amplitude of the response. However, (iii) an increase in the total incident flux will further ionize the disk so that it will reflect more efficiently. Therefore, an increased incident flux may also decrease the thermalized part (i.e., it will suppress the magnitude of the effects described above).  Figure~\ref{fig:mass} shows that the response amplitude increases with increasing BH mass, as expected.  But it does not increase exactly proportionally to $M_{\rm BH}$, due to the reason (iii) we mentioned above.

The response amplitude increases with decreasing accretion rate (Figure~\ref{fig:mdot}). This is caused by smaller temperature of the accretion disk as explained in point (ii) above when discussing the effect of BH mass. The response amplitude also increases with higher corona height and steeper $\Gamma$. As discussed in the previous section, the thermalized flux increases with increasing corona height and with the steepening of the power-law photon index of the X--ray corona. The inevitable consequence of this effect is an increase in the amplitude of the response function since more flux is emitted at a given wavelength (see Figures~\ref{fig:height} and \ref{fig:gamma}).

Finally, the response amplitude decreases with increasing inclination angle. This is a projection effect, since we see a smaller overall disk area as the inclination increases. Actually, at inclinations above $40^\circ$ the response starts to decay at much earlier times. This is due to the fact that the expanding reverberating part of the echo on the near side of the disk (closer to the observer) moves faster outward, outside of the disk region contributing to the given wavelength (see Figure~\ref{fig:theta}).

\subsection {The effects of BH spin}
For high BH spins, the disk inner edge moves closer to the BH and the disk temperature increases towards the center. However, this has little effect on the response functions. As we already discussed, the inner disk region contributes minimally to the response function (in the wavebands we study in this work). The emission from the inner disk affects the response function only at early times, so it does not affect substantially neither its width nor its amplitude. However, the BH spin affects the disk response in one more way. The spin determines the value of \mdot\ in physical units ($\rm M_\odot yr^{-1}$). The accretion rate in these units is higher for low spins (for the same $\dot{m}/\dot{m}_{\rm Edd}$), and the disk is hotter. Hence, for $a^\ast = 0$, the response functions are broader and have a lower amplitude compared to the response functions when $a^\ast = 1$. 

\subsection{The inner and outer disk radius effects}
\label{sec:Rinout}

The effects of changing the inner disk radius ($R_{\rm in}$) are shown in Figure~\ref{fig:Rin}. The response functions start at later times as $R_{\rm in}$ increases because the X-ray photons take more time to reach $R_{\rm in}$. At later times, when the outer disk elements contribute to the reverberation signal, the response functions are almost identical to the responses when $R_{\rm in}=R_{\rm ISCO}$. We note that the response amplitude increases (slightly) with larger $R_{\rm in}$ because $E_0$ changes with $R_{\rm in}$. Therefore, the incident X-ray spectrum and consequently the reverberated flux as well will be slightly different. However, the net effect is an overall decrease of the reprocessed flux as the inner disk radius becomes larger, as we miss an increasing fraction of the X-ray reverberation in the inner disk.

The opposite effect is observed when the outer disk radius ($R_{\rm out}$) decreases (Figure~\ref{fig:Rout}). The response functions in this case start at the same time (in all the wavebands). However, $\Psi$ becomes narrower as $R_{\rm out}$ decreases. This is because the outer disk (which is detected at later times) is missing in this case. This effect is more pronounced at longer wavelengths, due to the fact that the outer disk contributes mainly in the optical/near IR bands\footnote{In fact, the response functions in the {\it z'}-band appear to stop ``abruptly" in all cases we considered, mainly because the outer parts of the $R_{\rm out}=10^4 ~R_{\rm g}$ disk, for the parameters values we consider, are still quite ``warm" and they contribute a noticeable flux in the near-IR bands.}. As with $R_{\rm in}$, the net effect, mainly at longer wavelengths, is the overall decrease in the variable, reprocessed component. 

The change of the response width is more significant when $R_{\rm out}$ decreases. The start time of the response when $R_{\rm in}$ increases changes by a few hundredths of a day (Figure~\ref{fig:Rin}; we note that the time axis in all similar figures is logarithmic for clarity reasons), and is exactly the same in all bands. 
On the other hand, the changes in the width of the response when the outer disk radius decreases are of the order of many days, even in the UV bands (i.e in the $W2$ band, as shown by the second from left panels in Figure\,\ref{fig:Rout}). This will have a significant effect in the expected delays, which will be wavelength dependent.

\subsection{The dependence on X-ray luminosity}
\label{sec:Lx}

X-rays play a major role in thermal reverberation studies. Thus, it is important to understand how the response function depends on the X-ray luminosity. This will help us understand better the relation between X-ray and UV/optical variability in AGN. 

Figure~\ref{fig:Lx} shows $\Psi$ for many values of the observed, 2--10~keV band luminosity of the X-ray flash, $L^{\rm flash}_{\rm Xobs, Edd}$. We considered 2--10 keV luminosity values that may be large for AGN\footnote{According to Figure 10 in \cite{Lusso2012}, the (observed) 2--10 keV luminosity in AGN can be up to 0.1 of the Eddington luminosity.}, but these high values are necessary to reveal clearly the relation between the illuminating X--rays and the disk response. We recall that the disk response is already normalized to $L^{\rm flash}_{\rm Xobs, Edd}$, see Equation~(\ref{eq:psi}). So, if the response would simply scale linearly with $L^{\rm flash}_{\rm Xobs, Edd}$, we would expect the responses in Figure~\ref{fig:Lx} to overlap in all bands. However, this is not the case. This plot shows that the disk response depends on the incident X-ray flux in a non-linear way.  

The response functions (in all bands) decrease in amplitude and broaden in time as $L^{\rm flash}_{\rm Xobs, Edd}$ increases. This is mainly caused by the fact that the thermalized flux does not contribute to the response in each waveband in the same way at all times. As $L^{\rm flash}_{\rm Xobs, Edd}$ increases, the temperature of the disk elements also increases. {\em At early times}, the peak black body emission of the inner disk elements is at wavelengths which are shorter even than 1158~\AA. Therefore, we are in the Rayleigh-Jeans part of the BB spectrum emitted by these elements (at all wavebands). Consequently, the increase in flux (in each waveband) is not proportional to the increase in the incident X-ray flux. Therefore, when we divide the response by the illuminating $L^{\rm flash}_{\rm Xobs, Edd}$, the ratio decreases. {\em At later times}, it is the Wien part that contributes to the flux in each waveband. At the Wien part of the BB emission, the change in flux is larger than the change in the incident X-ray flux. So the increase in the response when the Wien part contributes is stronger than the increase in the $L^{\rm flash}_{\rm Xobs, Edd}$. Hence, when we normalize the response with $L^{\rm flash}_{\rm Xobs, Edd}$, the response is larger with increasing $L^{\rm flash}_{\rm Xobs, Edd}$, at later times. This effect explains the larger width of the responses with increasing $L^{\rm flash}_{\rm Xobs, Edd}$. For example, in the top rightmost panel of Figure~\ref{fig:Lx}, for the longest wavelength, the $z'$ band, the response does not reach the Wien part at the disk outer radius, $R_{\rm out}$, and thus the response does not increase with the X-ray luminosity even when it starts fading (the effect is visible, but tiny, in the case of $a^\ast = 1$ where the disk is colder).

There is yet another effect contributing to the response amplitude change with X-ray luminosity. This effect exhibits itself mainly at the beginning of the response when the inner parts of the disk reverberate. By increasing the X--ray luminosity, $F_{\rm inc}(R,t)$ increases. This leads to an increase in the ionization of the innermost parts of the disk and the fraction of the disk's reflection flux (i.e., $F_{\rm ref}(R,t)$). The reflection flux increases faster with the disk ionization than the incident flux due to higher disk reflectivity. Therefore, overall, the absorbed X-ray flux increases slower than the increase in $L^{\rm flash}_{\rm Xobs, Edd}$ and, when we plot the disk responses normalized to $L^{\rm flash}_{\rm Xobs, Edd}$, their amplitude will decrease as we increase the incident X-ray luminosity, at early times. At later times, we detect emission from the outer disk. At far enough distances from the center, the illuminating X-ray flux, which generally decreases as $\sim 1/R^2$, is always small enough that the disk is neutral and thus this effect dies out at larger response times. 

\begin{figure}
\centering
\includegraphics[width=0.98\linewidth]{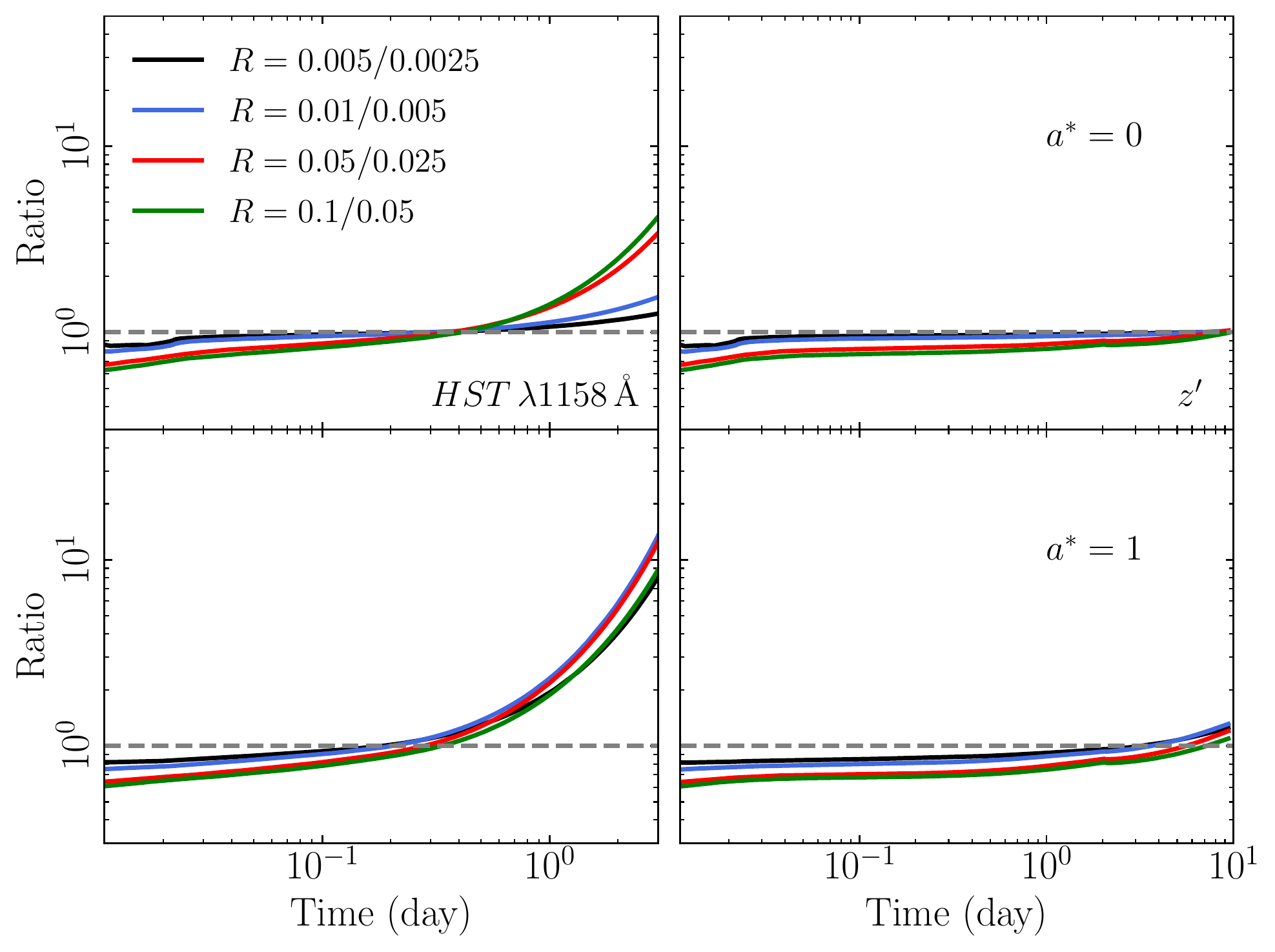}
\caption{ Ratio of response functions  in the $HST\lambda1158$ and $z'$ bands (left and right columns) and spins of 0 and 1 (upper and lower rows, respectively). See \S \ref{sec:Lxnonlin} for details.}
\label{fig:ratioLx}
\end{figure}

\subsubsection{Implications of the non-linear disk response}
\label{sec:Lxnonlin}

The non-linear disk response to the illuminating X--rays complicates the modelling of thermal reverberation in AGN. Equation~(\ref{eq:Fobs}) is valid, but if the disk response function depends on the incident X--rays, the convolution integral will be difficult to solve as $\Psi$ must change continuously, according to the variable input $L_{\rm X}$. This may be important when using Equation~(\ref{eq:Fobs}) and the observed X--rays to predict model UV/optical light curves (in all wavebands). In order to explore this issue, we used the response functions plotted in Fig.~\ref{fig:Lx} and we computed their ratio when the ratio of the incident X--ray luminosity is equal to 2. The results are plotted in Figure~\ref{fig:ratioLx}. 

The ratio of the responses is relatively flat in the optical band, both for spin 0 and 1 (right plots in Figure~\ref{fig:ratioLx}). The ratio is between $0.6-0.8$ at early times, because the response amplitude decreases when the X--ray luminosity increases (as we discussed above). There is a slight trend for the ratio to increase at later times (because the width of the response increases), but it is not very strong. 

The plots indicate that it may be possible to use Equation~(\ref{eq:Fobs}) with $\Psi$ equal to the disk response to the mean X--ray luminosity. For example, let us suppose that a source varies by a factor of 4 in the 2--10~keV band, say from 0.0025 to 0.01 (in Eddington units; black and blue curves in Figure~\ref{fig:ratioLx}), or from 0.025 to 0.1 (red and green curves in the same figure). Let us know assume that we use the response function for $L_{\rm Xobs,Edd}^{\rm flash}=0.005$ and 0.05, respectively, in Equation (\ref{eq:Fobs}). This X--ray luminosity is almost equal to the mean of the minimum and maximum values (in both cases). The use of a fixed response in Equation~(\ref{eq:Fobs}) means that we will be overestimating/underestimating the reverberating flux when the X--ray luminosity is larger/smaller than the mean. But, according to the black/blue and red/green lines in the right plots of Figure~\ref{fig:ratioLx}, we will over/under estimate the flux by almost the same factor each time. Therefore, the resulting reverberating flux should be approximately correct, on average. The left panels of Figure~\ref{fig:ratioLx} show the disk response functions in the far-UV band. Here, the change in the response width with incident X--ray flux are more pronounced at time scales longer than $\sim$ day. However, the black/blue and red/green lines are still (almost) identical. We therefore believe that the arguments above can still apply in this case.

We investigate in the Appendix~\ref{app:lightcurves} the non-linear disk response to X--rays in more detail. Our results show that, even if X--rays vary by a factor up to 10, we can use the disk response that corresponds to the mean X--ray luminosity in Equation~(\ref{eq:Fobs}), to estimate model UV/optical light curves, but also to model time-lags and reverberation fractions (see the following sections for the definition and modelling of these quantities).  
\section{Application to observations}
\label{sec:application}

The model disk responses can be used to study the X-ray-UV/optical correlation. One possibility is to use Equation~(\ref{eq:Fobs}) and the observed X-rays in order to produce model UV/optical light curves and compare them  with the observed ones. This would be the best way to test the validity of the model assumptions and to constrain the disk/corona geometry. However, this is a formidable task. 

The sampling rate of the X--ray light curves is crucial for the proper estimation of the X-ray reprocessed UV/optical flux. If the average cadence of the X-ray light curve is larger than the time scale when the bulk of the UV/optical flux is emitted, interpolation of the X-ray light curve will be necessary. For example, Figure~\ref{fig:mass} shows that the UVW2 response has decreased to 10\% of its maximum amplitude after $\sim 0.6, 1.5$ and 3 days, in the case of a $10^7, 5\times 10^7$ and $10^8$ M$_{\odot}$. Therefore, we need an X--ray sampling rate at least ten times smaller, roughly speaking, in order to be able to compute the integral in Equation~(\ref{eq:Fobs}) correctly, for this band. We would need an even denser sampling if we wish to compute the model flux at shorter wavelengths (the constrain is more relaxed of course at longer wavelengths). This would of course also depend on the X-ray variability amplitude, but it is possible that  the resulting UV/optical light curves may tell more about the way the interpolation was done than the physical properties of the source. 

Another way to compare the model predictions with data is through power-spectrum (PSD) analysis. Equation\,\ref{eq:Fobs} predicts a unique relation between the X-ray and the UV/optical PSDs \citep[see e.g.,][]{Panagiotou20}. The most common and more frequent method to study the X-ray/UV and optical correlation is cross-correlation analysis of multi-wavelength, monitoring observations. Cross-correlation techniques are used to estimate the time-lags between the observed X-ray, UV and optical light curves, as a function of the wavelength. We used the model disk transfer functions presented in the previous Section and we computed the model time-lag spectra for all the model parameters we consider in this work. We present our results below. 

\subsection{Cross correlation and time lags}

The response function determines, to a large extent, the cross-correlation between the X-ray and the UV/optical light curves. It is commonly accepted that the centroid of the response function, \tcen,  is representative of the maximum peak in the cross-correlation function (CCF) that is measured between the X-ray and the UV/optical light curves. The centroid is defined as follows:
\begin{equation}
\tau_{\rm cen} (\lambda_{\rm c})  = \frac{\int t\,\Psi(t, \lambda_{\rm c})\,{\rm d}t}{\int \Psi(t, \lambda_{\rm c})\,{\rm d}t}.
    \label{eq:meantau}
\end{equation}

\noindent We note that this is an approximate measure of the CCF peak, because the CCF is the convolution of the transfer function and the continuum auto-correlation function \citep[e.g., Equation (16) in][]{Peterson93},

\begin{equation}
    {\rm CCF}(\tau) = \int_{-\infty}^{+\infty} \Psi(t)\,{\rm ACF_{X}}(\tau - t)\,{\rm d}t,
    \label{eq:CCF}
\end{equation}
\noindent where $\rm ACF_{X}$ is the auto-correlation function of the X-ray light curve. The X-ray PSDs have a power-law like shape, that extends to very low frequencies with a slope of $\sim -1$ \citep[e.g.,][for NGC~4051]{Mchardy04}. This implies a long memory, hence a broad X-ray ACF.  Consequently, the X-ray ACF should affect the value of the time lag between the X-ray and the UV/optical light curves. However, the X-ray ACF is common to the CCF between X--rays and any UV/optical band, and it should affect all CCFs in the same way, i.e. we would expect the X-ray ACF to causing the same systematic effect to all CCFs. Therefore, we believe that the X-ray ACF may affect less the time-lags between a reference band in the UV, and any other UV/optical light curve. We would expect the time-lags within the UV/optical light curves to depend, mainly, on the disk response function, and not on the X-ray ACF.

We used Equation~(\ref{eq:meantau}) and the model disk responses to compute model time lags spectra, for all the model parameters. The respective plots are shown in the Appendix (Figures~\ref{fig:tau_mdot}-\ref{fig:tau_Lx}). In general, \tcen\ increases with increasing wavelength, because the width of the disk response increases with the wavelength (for all parameters).

 The slope of the $\tau_{\rm cent}$ vs $\lambda$ relation depends on the corona height (see Figure\ref{fig:tau_height}). When the corona is located close to the BH, the slope of the time-lags spectrum is $\sim 1.34$, i.e., equal to the ``canonical" slope of 4/3 that is expected in the case of a point-like corona located close to the BH, illuminating a NT disk. As the corona height increases, the time lags increase at all wavelengths. But, they increase more at shorter wavelengths, because the width of the response increases more at shorter wavelengths. As a result, the slope of the time-lag spectra flattens. For example, when $h=100$ R$_{\rm g}$, the slope is $\sim 1.1$. 

The time lags increase with increasing BH mass, accretion rate and X-ray luminosity. These three parameters affect the normalization of the $\tau$ vs $\lambda$ relations, but not the slope, because they affect the width of the transfer functions by the same factor, at all wavelengths. The time-lag spectra increase as the spectral slope $\Gamma$ steepens, because the width of the response function increases accordingly (Section~\ref{sec:Psiwidth}). Interestingly, the time-lag spectra do not depend on the inclination. As the inclination increases, the response start time decreases, the width increases, and the response amplitude decreases (see Figure~\ref{fig:theta}). But, apparently, all these changes happen in such a way that the centroid of the response is not affected. It is as if the time lags are set by the physical properties of the system, and do not depend on the viewing angle.  

The time-lag spectra do not depend significantly on $R_{\rm in}$. 
As we discussed in Section~\ref{sec:Rinout}, as $R_{\rm in}$ increases, the rise time of the disk response increases. The effect is more pronounced in the UV bands. This shift in the rise time leads to an increase in the centroid of the response function (and hence of $\tau$), as seen in Figure~\ref{fig:tau_Rin}. But the effect is subtle, and can be seen only when the inner radius is at least 100~$R_{\rm g}$. On the other hand, $R_{\rm out}$ has a profound effect on the time-lags (see Figure~\ref{fig:tau_Rout}). The time-lag spectra flatten significantly as the outer radius decreases. In fact, $R_{\rm out}$ is the only parameter that can cause a flattening of the time-lag spectra even at wavelengths shorter than $5000$~\AA. This is due to the fact that the width of the disk response is significantly shortened, even at short wavelengths, as $R_{\rm out}$ decreases. Figure~\ref{fig:tau_Rout} shows that time-lag spectra can be very flat if the outer disk radius is smaller than 5000~$R_{\rm g}$. Observed time-lag spectral slopes of the order of $\sim 0.5$ \citep[like NGC~4593, see][]{Mchardy18,Edelson19}, can be explained by the model, but only if $R_{\rm out}$ is smaller than $\sim 5000 ~R_{\rm g}$.

\subsection{An analytic prescription for the time-lag spectra}

We fitted all the model time-lag spectra, and we found that the following equation describes our results well: 

\begin{equation}
\tau_{\rm cen} (\lambda) = A(h_{10}) M_7^{0.7} f_1(\dot{m}_{0.05})f_2(L_{\rm X,0.01}) \lambda_{1950}^{B(h_{10})}~\rm day,
\label{timelagseq}
\end{equation}


\noindent where $\lambda_{1950}$ is the wavelength normalized to 1950~\AA, $h_{10}$ is the height of the lamp-post in units of 10~\rg, $M_7$ is the BH mass in units of $10^7 M_\odot$, $\dot{m}_{0.05}$ is the accretion rate in units of 5$\%$ of the Eddington limit, $L_{\rm X,0.01}$ is the observed, 2--10~keV luminosity in units of 0.01 of the Eddington luminosity, and,

\begin{eqnarray}
A(h_{10}) &=& 0.164 + 0.039  h_{10} - 0.0012  h_{10}^2 \\
B(h_{10}) &=& 1.346 - 0.037 h_{10} + 0.0013   h_{10}^2 \\ \label{eq:mdot_a1}
f_1(\dot{m}_{0.05}) &=& 0.823 + 0.193 \dot{m}_{0.05} - 0.023 \dot{m}_{0.05}^2 \\
 &+& 0.0012 \dot{m}_{0.05}^3 \nonumber \\
f_2(L_{\rm X,0.01}) &=& L_{\rm X,0.01}^{0.025} \left[ \frac{1}{2} \left( 1 + L_{\rm X,0.01} ^{0.79}\right) \right]^{0.38}
\end{eqnarray}
\noindent in the case when $a^{\ast}=1$ and, 

\begin{eqnarray}
A(h_{10}) &=& 0.27 + 0.031 h_{10} - 0.0007  h_{10}^2 \\
B(h_{10}) &=& 1.395 - 0.043  h_{10} + 0.0017   h_{10}^2 \\ \label{eq:mdot_a0}
f_1(\dot{m}_{0.05}) &=& 0.636 + 0.4 \dot{m}_{0.05} - 0.059 \dot{m}_{0.05}^2 \\ 
 &+& 0.0033 \dot{m}_{0.05}^3 \nonumber \\
f_2(L_{\rm X,0.01}) &=& L_{\rm X,0.01}^{-0.763} \left[ \frac{1}{2} \left( 1 + L_{\rm X,0.01} ^{0.12}\right) \right]^{14.03}
\end{eqnarray}
\noindent in the case when $a^{\ast}=0$.

These equations are valid for the time-lag spectra when $\lambda \le 5000$~\AA~ because the time-lag spectra flatten at longer wavelengths for certain combinations of the model parameters. In all these cases, the disk is hot enough to contribute to the disk response at long wavelengths, even at radii larger than 10$^4 ~R_{\rm g}$. Since this is the largest $R_{\rm out}$ radius we considered, the responses at these wavelengths show an abrupt cut-off at long time scales, and the respective time-lags are smaller than the values if we had considered a larger outer disk radius.  

Equation~(\ref{timelagseq}) shows that the time-lags depend on many parameters, in a non-linear way. Of course one can always fit a simple power-law function to the observed data, but the use of Equation~(\ref{timelagseq}) to fit observed time-lag spectra could be more beneficial. In fact, as long as a BH mass estimate is available, and an average, observed 2--10~keV X-ray luminosity can be determined by the X-ray observations of a source, it is straightforward to use Equation~(\ref{timelagseq}) to fit the observed time-lags, and determine the $(\dot{m}_{\rm Edd}, h)$ values that could fit the data best.


\subsection{Reverberation fraction}
\label{sec:Revfrac}

Although the reverberation signal has been seen in a few sources so far, some AGN, which are very variable in X-rays, do not show any correlation between X-rays and the UV/optical \citep[see e.g.,][]{Pawar17,Buisson18, Morales19}. This could be either due to the fact that thermal reverberation does not take place in these sources, or it is too weak to be detected. 

To investigate this issue, we defined the `reverberation fraction', $\cal{R}_{\rm rev}$, as the ratio between the average reverberated flux and the total (reverberated plus NT) disk flux in a given waveband, as follows

\begin{equation}
    {\cal{R}}_{\rm rev}(\lambda_{\rm c}) =  \frac{L_{\rm X,Edd} \int_{0}^{\infty} \Psi (\lambda_{\rm c}, t)\,{\rm d}t}{F_{\rm NT}(\lambda_{\rm c}) + L_{\rm X,Edd} \int_{0}^{\infty} \Psi (\lambda_{\rm c}, t)\,{\rm d}t}.
\end{equation}
\noindent We computed $\cal{R}_{\rm rev}$ for all the model parameters, and we present the results in Figures~\ref{fig:frev_mass}-\ref{fig:frev_Lx} in the Appendix. 

The reverberation fraction depends on the incident X-ray flux, the amplitude of the response function and the temperature of the accretion disk. In some cases, this ratio is close to one, which means that the flux in these wavebands is dominated by the flux generated by X-ray absorption. Objects with physical parameters which predict large $\cal{R}_{\rm rev}$ will be ideal targets for the study of the X-ray and UV/optical correlation.

A small decrease in \Rfrac\ can be seen with increasing $\lambda$, in almost all cases. In addition, \Rfrac\ is larger for a rotating BH compared to the Schwarzschild case. This is due to the fact that for the same \mdot\ in Eddington units, the accretion disk is hotter for lower spins, hence $F_{\rm NT}(\lambda_{\rm c})$ is larger, and the reverberation signal is smaller. It appears then that it is better to search for the reverberation signal at shorter wavelengths, in maximally rotating BH.

For a fixed BH mass, the UV/optical flux due to X-ray heating (thus \Rfrac\ as well) increases with increasing X-ray flux, increasing corona height (because the response amplitude increases with height, as we explained in Section~\ref{sec:Psiamplitude}) and decreasing accretion rate (because the disk becomes colder). The reverberation fraction also increases as the spectrum steepens, because the X-ray flux absorbed by the disk also increases in this case. For fixed \Lx\ and \mdot\ (in units of Eddington), the amplitude of the response functions increases with increasing BH mass (Figure~\ref{fig:mass}). At the same time, the flux of the NT disk increases as well. As a result, \Rfrac\ increases with mass, albeit with a moderate rate (Figure~\ref{fig:frev_mass}). 

The inner and outer disk radii also affect the reverberation factor, although not significantly. 
The strength of the reverberation component at optical wavebands decreases, as $R_{\rm out}$ becomes smaller (Figure~\ref{fig:frev_Rout}). This is because the response function becomes narrower with decreasing $R_{\rm out}$, so its integral over time decreases accordingly. The reverberation fraction increases when $R_{\rm in}$ gets larger (Figure~\ref{fig:frev_Rin}). This is because the disk flux decreases by considering larger values of $R_{\rm in}$. However, the effect is not very strong.

The model predicted reverberation fraction can be used to explain the results from monitoring campaigns of some X-ray bright, highly variable NLS1s, such as IRAS~13224--3809 \citep{Buisson18}, Mrk~817 \citep{Morales19}, and 1H 0707--495 \citep{Robertson15,Pawar17}. The data showed very weak, or even absent, UV/optical variability while the X-rays were highly variable. 

According to our model, absence of a significant correlation between X-ray and UV/optical could be explained on the case when the reverberation fraction is small. This is to be expected in the case of a low corona height and high accretion rate. For example, according to  Figure~\ref{fig:frev_height}, \Rfrac$\lesssim 0.55$ and $\lesssim 0.2$ in the case of spin 1 and 0, respectively, when $h \le 5 R_{\rm g}$ and  the accretion rate is 5 per cent of the Eddington limit (the fiducial value). However, if the accretion rate increases to 50~\% of $\dot{m}_{\rm Edd}$, the reverberation fraction will decrease to $\sim 0.15$ and 0.02 of the total flux (in the far-UV, slightly less in the other wavebands) in the case of spin 1 and 0, respectively. In this case, it will be difficult to detect significant variations in the total observed flux. 


\begin{table}
\centering
\caption{A qualitative summary of the effects of each of the parameters studied in this work on the response function amplitude, start time, and width, and the time lag and reverberation fraction.}
\begin{tabular}{llllll}
\hline \hline
Parameter 	&	\multicolumn{3}{c}{$\Psi$}	&	$\tau_{\rm cen}$	&	$\mathcal{R}_{\rm rev}$	\\ 
 	&	amplitude	&	 start time	&	width	&	&\\\hline
$a^\ast \nearrow$	&	$\nearrow$	&	$-$	&	$\searrow$	&	$\searrow$	&	$\nearrow$	\\
$M_{\rm BH} \nearrow$	&	$\nearrow$	&	$\nearrow$	&	$\nearrow$	&	$\nearrow$	&	$\nearrow$	\\
$\dot{m}/\dot{m}_{\rm Edd} \nearrow$	&	$\searrow$	&	$-$	&	$\nearrow$	&	$\nearrow$	&	$\searrow$	\\
$h \nearrow$	&	$\nearrow$	&	$\nearrow$	&	$\nearrow$	&	$\nearrow$	&	$\nearrow$	\\
$\Gamma  \nearrow$ 	&	$\nearrow$	&	$-$	&	$\nearrow$	&	$\nearrow$	&	$\nearrow$	\\
$\theta  \nearrow$	&	$\searrow$	&	$\searrow$	&	$\nearrow$	&	$-$	&	$-$	\\
$R_{\rm in} \nearrow$	&	$\searrow$	&	$\nearrow$	&	$\searrow$	&	$\nearrow$	&	$\nearrow$	\\
$R_{\rm out} \nearrow$	&	$-$	&	$-$	&	$\nearrow$	&	$\nearrow$	&	$\nearrow$	\\
$L_{\rm Xobs,Edd} \nearrow $	&	$\searrow$	&	$-$	&	$\nearrow$	&	$\nearrow$	&	$\nearrow$	\\
\hline \hline
\end{tabular}
\label{table:summary}
\end{table}

\section{Summary and discussion}
\label{sec:discussion}

We presented the results from a detailed analysis of thermal reverberation in the UV/optical bands caused by the X-ray illumination of the disk by a point-like X-ray source in AGN. We considered a number of possible physical parameters that can affect the disk response to X-ray illumination. Main inputs to our code are the {\it observed} 2--10~keV X-ray luminosity (in Eddington units), spectral slope ($\Gamma$), and the observed high energy cut-off ($E_{\rm cut,obs}$). We considered all the appropriate GR effects and we determine: 1) the {\it intrinsic} X-ray luminosity and high energy cut-of, $E_{\rm cut,intr}$, as well as the low-energy cut-off, $E_0$, of the corona (using the respective observed values, $M_{\rm BH}$, $\dot{m}$, and the height $h$), and 2) the incident X-ray spectrum at each disk element (in its rest frame). The model then computes the appropriate reflection spectrum, taking into consideration the ionization state of each disk element. In this way, the model accurately computes the incident and the reflected flux, hence the flux that the disk absorbs. We compared our model disk responses with responses that we computed ignoring the GR and/or the disk ionization effects (as it is frequently the case in past studies). We found that ignoring those effects can significantly alter the disk response.

We computed disk response functions in various wavebands. Table~\ref{table:summary} lists all the model parameters and how they affect the main characteristics of the disk response functions (i.e., amplitude, start time, and width). Our results show that the disk response to the corona luminosity is non-linear, due to the non-linear response of the disk thermalization to the variable X-ray illumination. 
We also investigated the effect of the inner and outer disk radius on the response functions. An inner disk radius that is larger than the ISCO radius may be relevant in cases where an extended warm corona with a large optical depth may be present on top of the inner disk \citep[see e.g.,][and references therein]{Mehdipour11, Petrucci13, Mehdipour15, Mehdipour18, Porquet18,Petrucci20, Matzeu20}. In that case, due to the high optical depth, all the  UV/optical photons from the innermost regions of the accretion disk will be shifted to soft X-rays, and the reverberation signal will be observed from disk radii that are larger than the radius of the warm corona. The main effect of increasing $R_{\rm in}$ is the increase in the rise time of the response, which implies that the amplitude of the reverberating component will decrease (this may be important when predicting the reverberating signal, especially in the UV bands). 

We used the response functions and we computed model time lags and the reverberation fraction \Rfrac\ (i.e., the fraction of reverberated flux over the total flux). This ratio can be used to explain the lack of UV/optical variability in the case of a few AGN which are highly variable in X-rays. Our results indicate that in the case of low heights and high accretion rates, the reverberating signal can have a very small amplitude that will make it difficult to detect and to study its variability. 

The time lags are affected by all the physical parameters that influence the disk response as well. Table \ref{table:summary} summarizes the effects of the model parameters to the time lags and \Rfrac. In some cases, the model time-lags spectra flatten at longer wavelengths, but this is due to the fact that we have considered an outer disk radius of $10^4~R_{\rm g}$. A larger $R_{\rm out}$ will eliminate this effect. However we note that, very flat time-lag spectra can be explained within the context of X-ray irradiation of accretion disks, but only if $R_{\rm out}$ is $\sim 1000~R_{\rm g}$ (or smaller). This possibility implies that time-lag spectra can impose constraints and in fact, in some cases, they may be used to measure the size of the accretion disk.

The model time-lag spectra we have computed can be used to fit observed time-lag spectra, however, there is a simpler way for this. We derived an analytic expression for the time-lags as a function of wavelength. Equation~(\ref{timelagseq}) can be used to fit the observed time-lag spectra at wavelengths shorter than $\sim 5000$ \AA. This will provide best-fit values for important physical parameters of the system, like the accretion rate and the corona height, as long as an accurate BH mass estimate is available. The luminosity term in this equation refers to the average, observed 2--10~keV luminosity, so it should be easily determined by the observations. 

Finally, our results indicate that any variability in the system will be imprinted in the time-lag spectra. In particular, long-timescale changes of the height of the corona, the observed X-ray luminosity, or the disk accretion rate, for a given source, will change the amplitude of the time-lag spectrum (and its slope, in the case of a variable height). These changes can be revealed through multi-epoch monitoring of bright sources.


\begin{acknowledgements}

MCG and MD acknowledge MEYS of Czech Republic for the support through the 18-00533S research project. MCG acknowledges funding from ESA through a partnership with IAA-CSIC (Spain). MD thanks his home institution, ASU, supported by the project RVO:67985815. ESK and IEP would like to thank ASU for hospitality. IEP acknowledges support of the International Space Science Institute, Bern. The figures were generated using Matplotlib \citep{Hunter07}, a {\tt PYTHON} library for publication of quality graphics.

\end{acknowledgements}

\software{Matplotlib \citep{Hunter07}, SciPy \citep{scipy}.}

\section*{Data availability}

The {\tt KYNXILREV} code used to compute the response functions is publicly available at \url{https://projects.asu.cas.cz/stronggravity/kynreverb/}. All the response functions and data presented in this paper are available upon request.

	\bibliographystyle{aasjournal}
	\bibliography{references}
\pagebreak

\appendix

\section{The non-linear disk response to X--rays.}\label{app:lightcurves}

To study the effects of the non-linear disk response to the incident X--rays, we performed the following experiment. We considered an X-ray light curve with a max-to-min ratio of 5 (the `low-var' case), and a light curve with a maximum variability amplitude of 10 (the `high-var' case). We used the fiducial parameters listed in Table~\ref{table:param}, and we assumed a maximally spinning black hole. The light curves were 8.5 days long, with $\Delta t=\rm T_g$. 

We constructed one thousand light curves in each case. The `low' and the `high-var' light curves are a random series of the $L_{\rm Xobs,Edd}^{\rm flash}$ values of 0.001, 0.0025 or 0.005, and 0.001, 0.005 or 0.1, respectively. Each value is chosen randomly, with equal probability. For each light curve, we used Equation~(\ref{eq:Fobs}) and we estimated the HST$\rm \lambda 1158$ and $g'-$band model reverberation light curves twice: in the first case we considered the exact response function for each $L_{\rm Xobs,Edd}^{\rm flash}$, while in the second time we assumed the response of the `intermediate' X-ray luminosity, being 0.0025 and 0.005, respectively. 

\begin{figure*}
\centering
\includegraphics[width=0.95\linewidth]{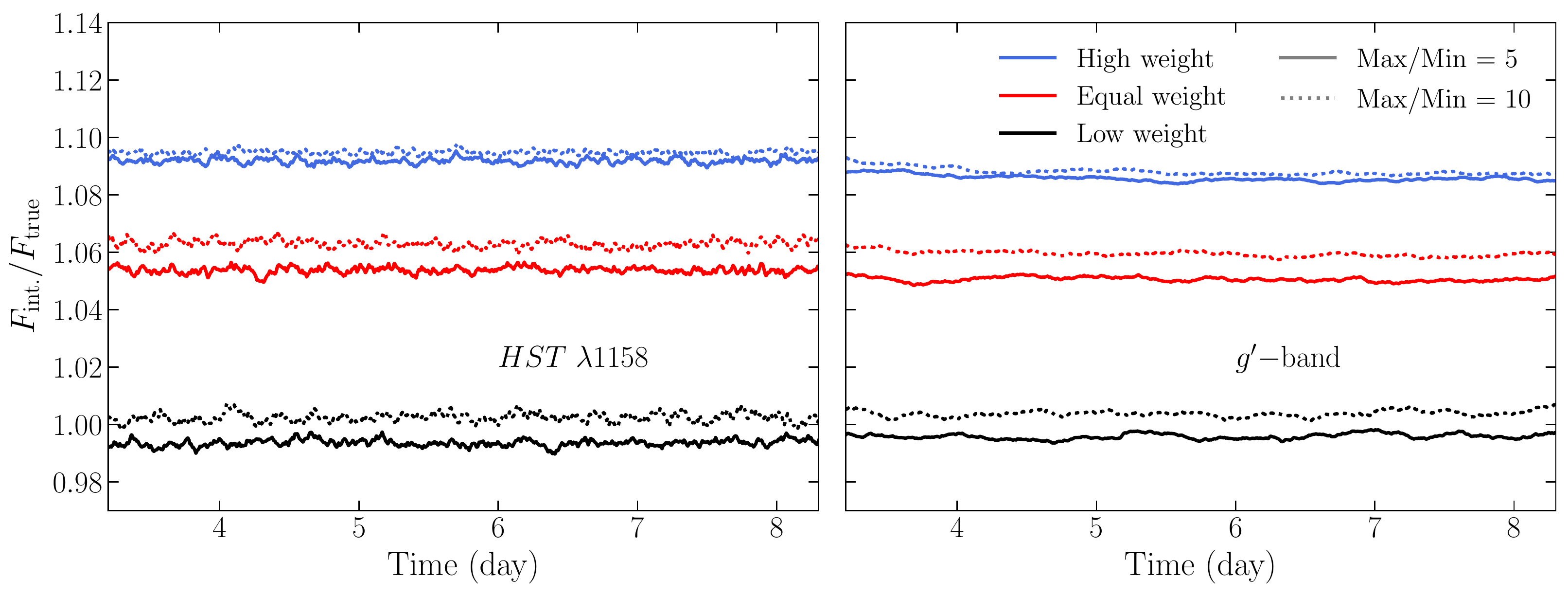}
\caption{The ratio of the reverberation light curves obtained by assuming the disk response corresponding to the intermediate X-ray luminosity over the correct response function for a max-to-min X-ray variability of 5 and 10 (solid and dotted lines, respectively). We assume that the source is most of the time in its lowest flux state, highest flux state or with an equal probability in all flux states (black, blue and red lines, respectively). The ratios are shown in the HST$\rm \lambda 1158$ and $g'-$ bands (left and right, respectively). See the Appendix~\ref{app:lightcurves} for more details.}
\label{fig:lightcurve}
\end{figure*}

The solid and dotted red lines in Figure~\ref{fig:lightcurve} show the mean ratio of the light curves obtained by assuming the response function of the intermediate X-ray luminosity over the light curves computed by using the exact response functions, in the `low-var' and `high-var' case, respectively. We show the ratio three days after the first X-ray flash, when the UV/optical flux has reached an equilibrium state. The results show that the difference between the model predictions when we use the exact responses and when we use the response for the intermediate X--ray flux, differ by $\sim 5-6$ per cent, even if the X--rays vary by a factor of 10. 

We repeated the same test by assigning different weights on $L_{\rm Xobs,Edd}^{\rm flash}$. We considered two scenarios where the X-ray source is in a `faint' or a `bright-state' for most of the time. To do so, the three $L_{\rm Xobs,Edd}^{\rm flash}$ values we mentioned above were not chosen randomly any more, but with a probability of  $[60\%, 30\%, 10\%]$ and $[10\%, 30\%, 60\%]$, respectively. We constructed again 1000 light curves, and we computed the mean ratio of the model predicted reverberation flux when we consider the exact responses and the response of the mean X--ray flux. The results are shown in the same figure (black and blue lines for the bright and the faint-state, respectively; solid and dotted lines for the `low' and `high-var' light curves, respectively). We note that the scatter of the 1000 simulated ratios around the plotted (average) ones is in the order of $0.6-1\%$ and $0.2-0.4\%$ only for the HST$\rm \lambda 1158$ and $g'-$band, respectively.

Our results show that, regardless of the X-ray variability amplitude, by assuming the response for the intermediate X-ray luminosity, we over-predict the reverberation signal in the far-UV and optical bands by no more than $\sim 9\%$, if the source is in a bright-state most of the time. If the source is in a faint-state for most of the time, the model reverberation flux will differ by less than $\sim 1$\% of the exact model value when we assume the disk response to the intermediate X--ray flux.

\section{Plots}

\begin{figure*}
\centering
\includegraphics[width=0.95\linewidth]{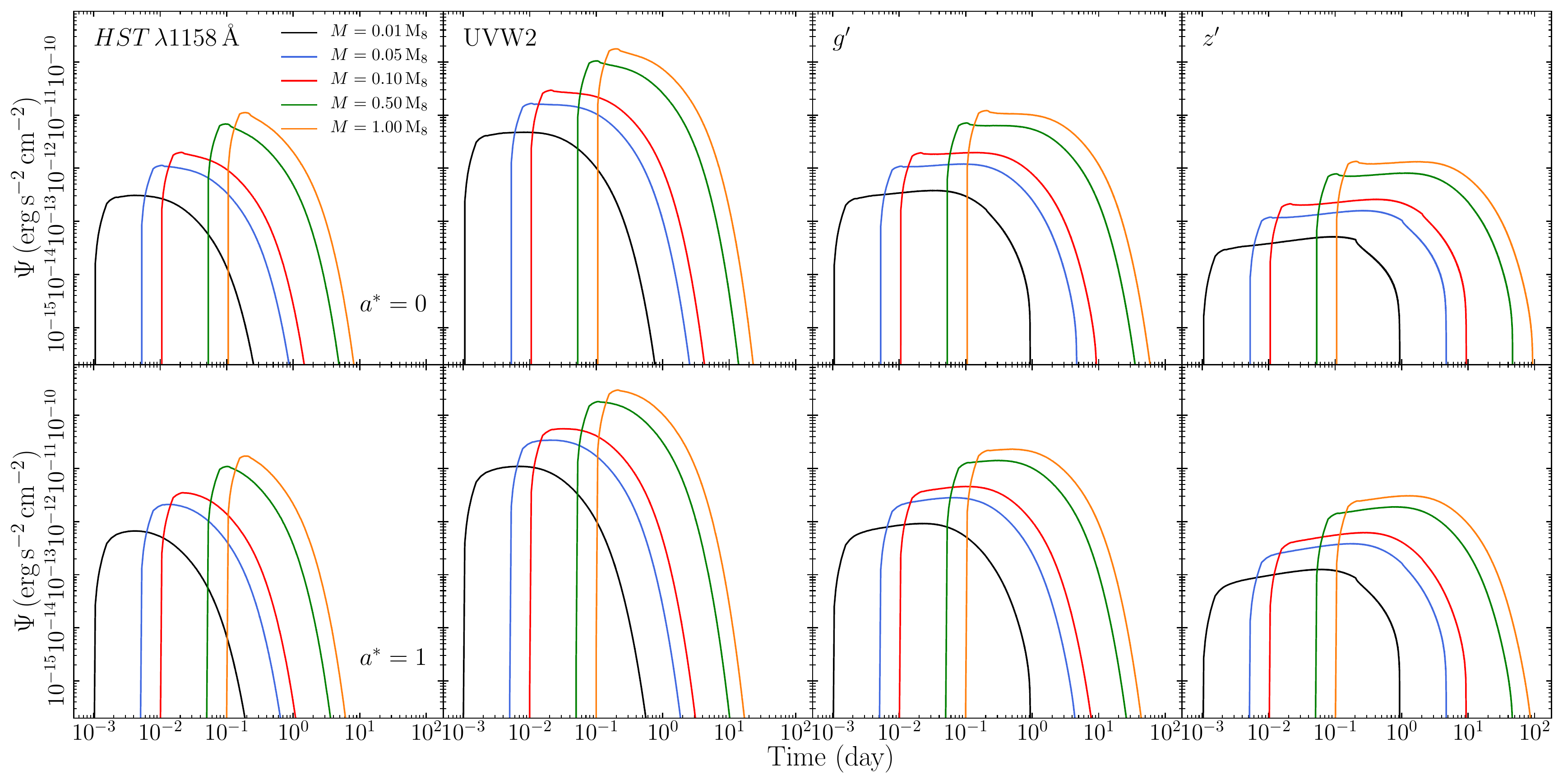}
\caption{Response functions for all BH masses we consider in this work (see Table~\ref{table:param}, in four different wavebands considering spins of 0 and 1 (top and bottom panels, respectively).}
\label{fig:mass}
\end{figure*}

\begin{figure*}
\centering
\includegraphics[width=0.95\linewidth]{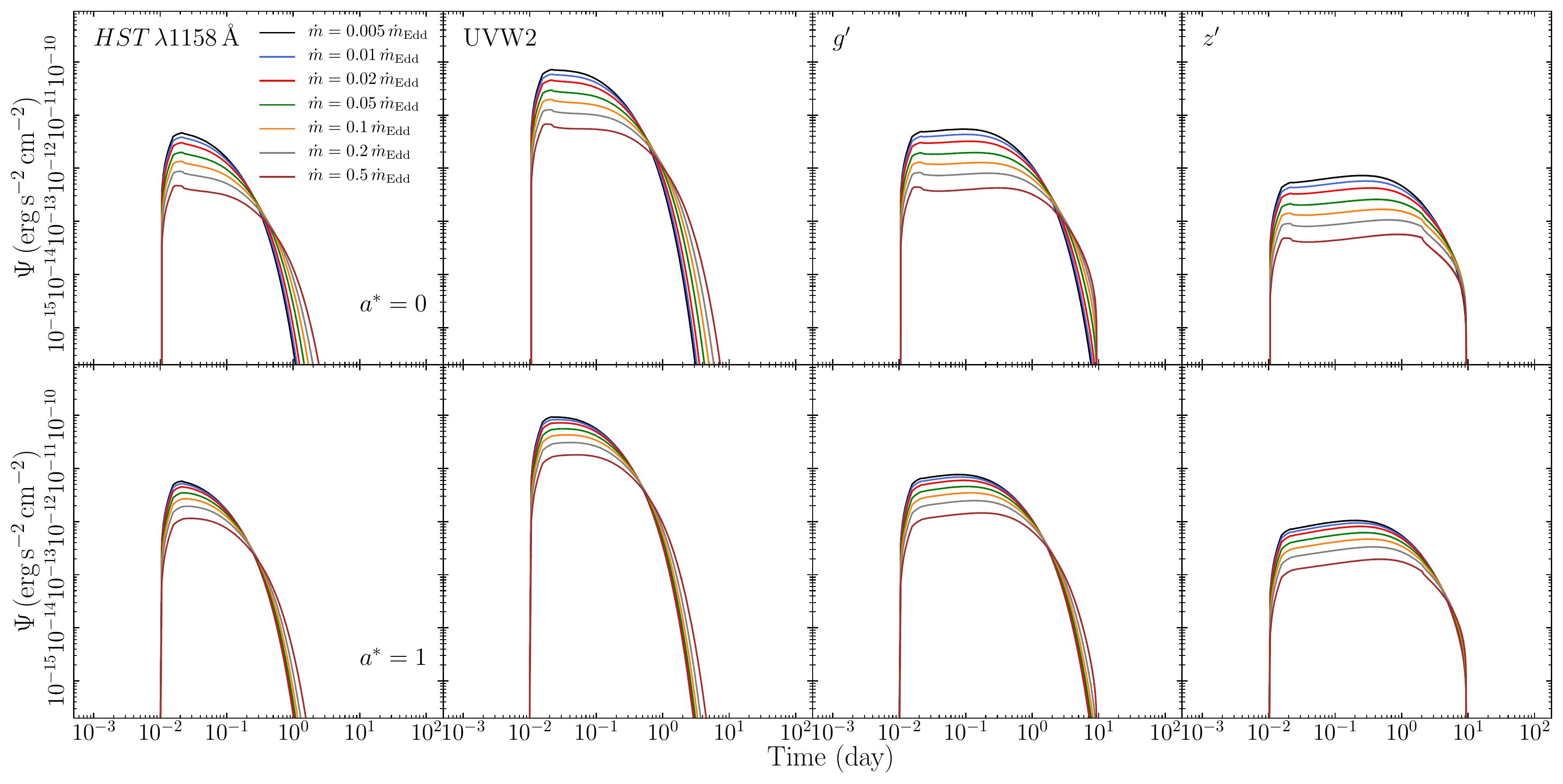}
\caption{Same as Figure \ref{fig:mass}, for the accretion rate.}
\label{fig:mdot}
\end{figure*}

\begin{figure*}
\centering
\includegraphics[width=0.95\linewidth]{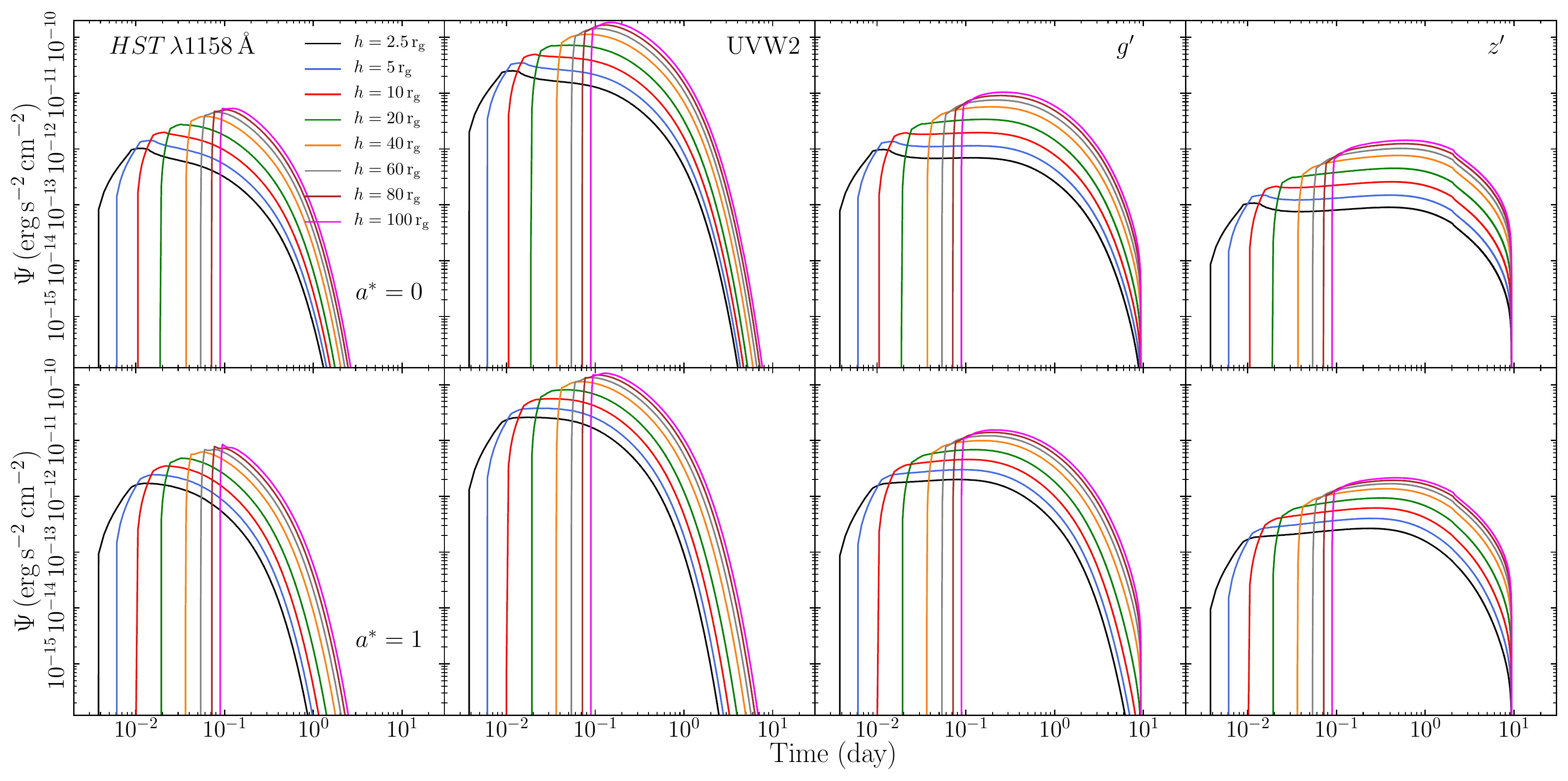}
\caption{Same as Figure \ref{fig:mass}, for the lamp-post height.}
\label{fig:height}
\end{figure*}

\begin{figure*}
\centering
\includegraphics[width=0.95\linewidth]{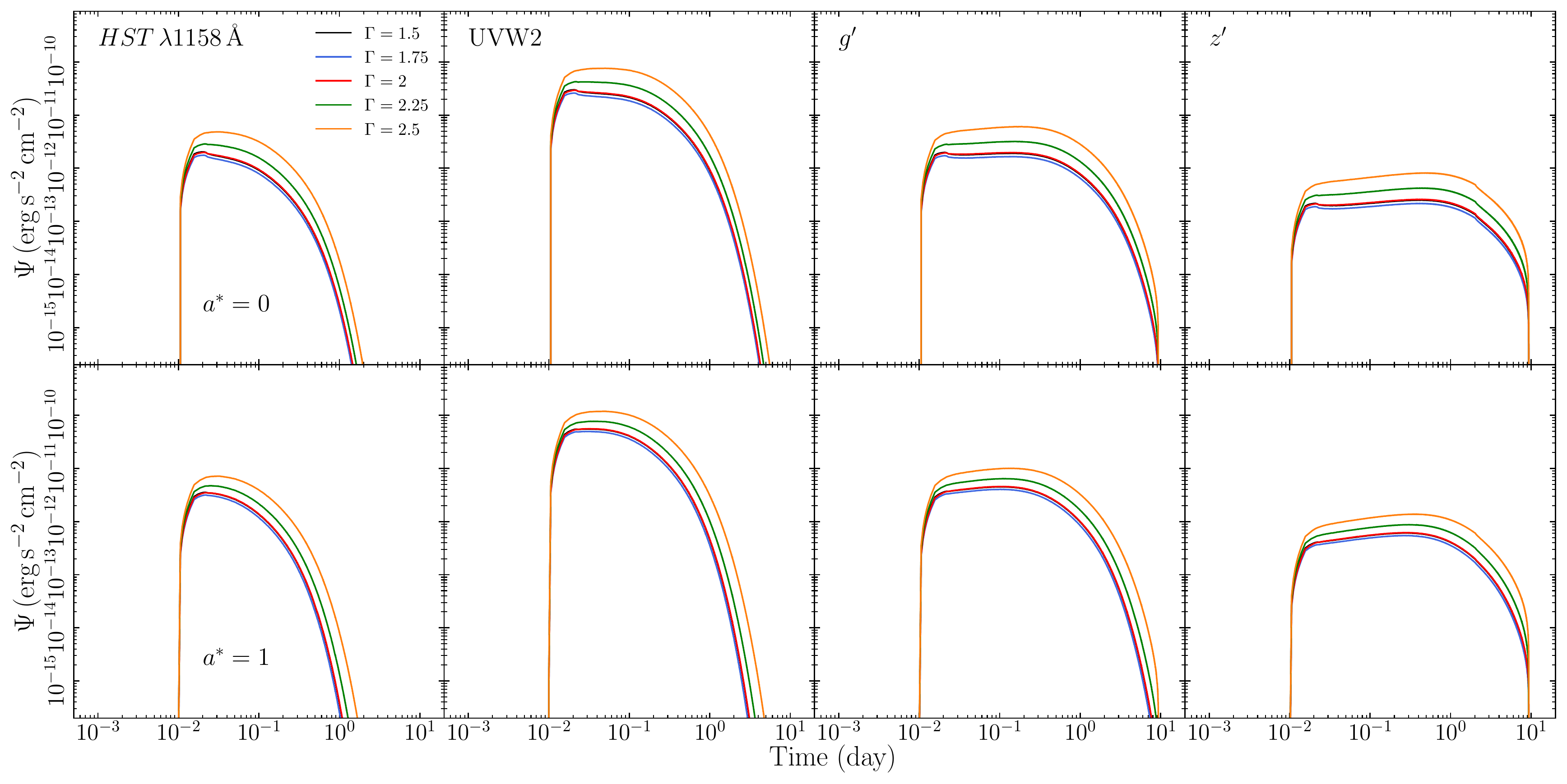}
\caption{Same as Figure \ref{fig:mass}, for the photon index.}
\label{fig:gamma}
\end{figure*}

\begin{figure*}
\centering
\includegraphics[width=0.95\linewidth]{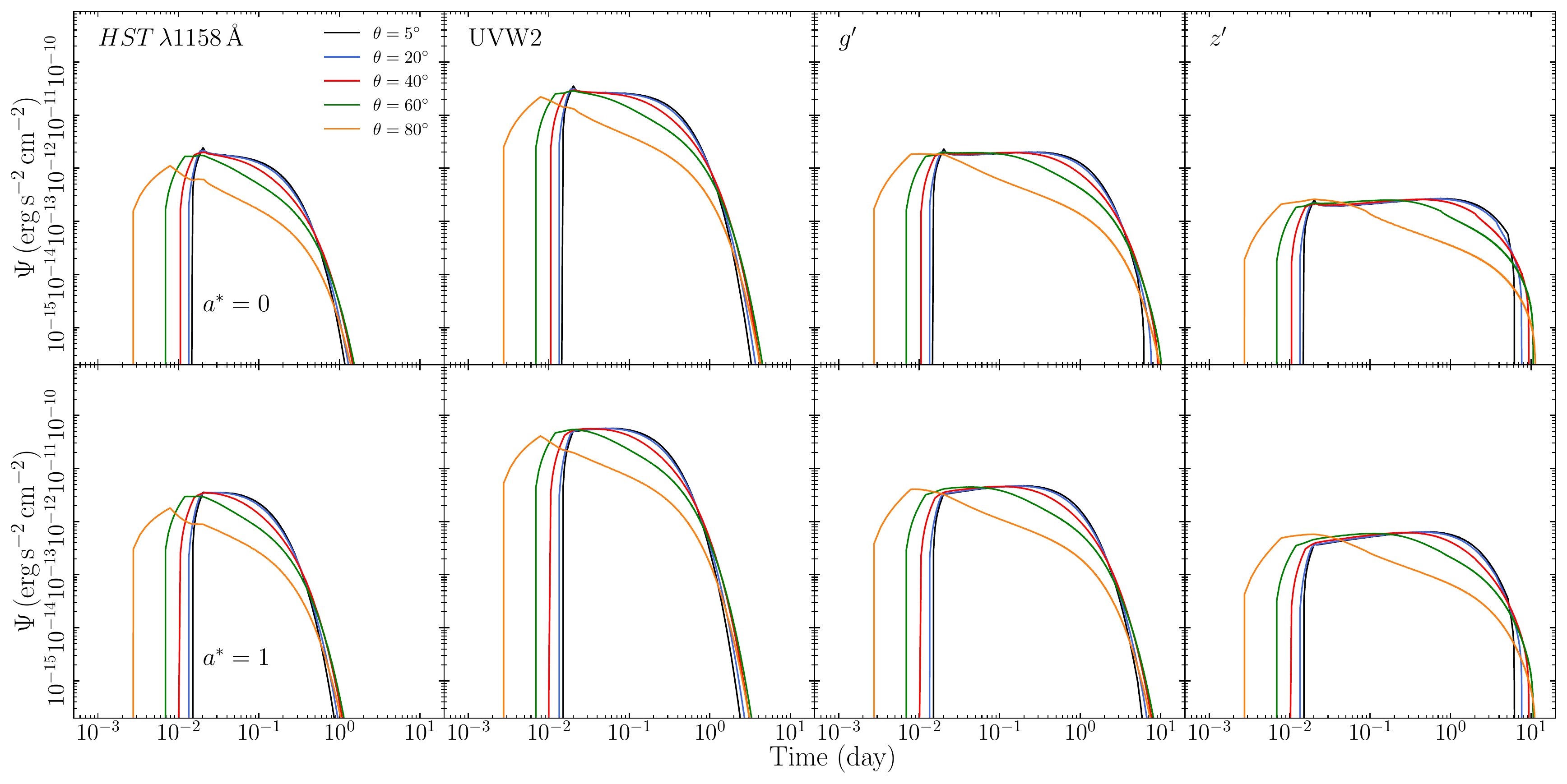}
\caption{Same as Figure \ref{fig:mass}, for the inclination angle.}
\label{fig:theta}
\end{figure*}

\begin{figure*}
\centering
\includegraphics[width=0.95\linewidth]{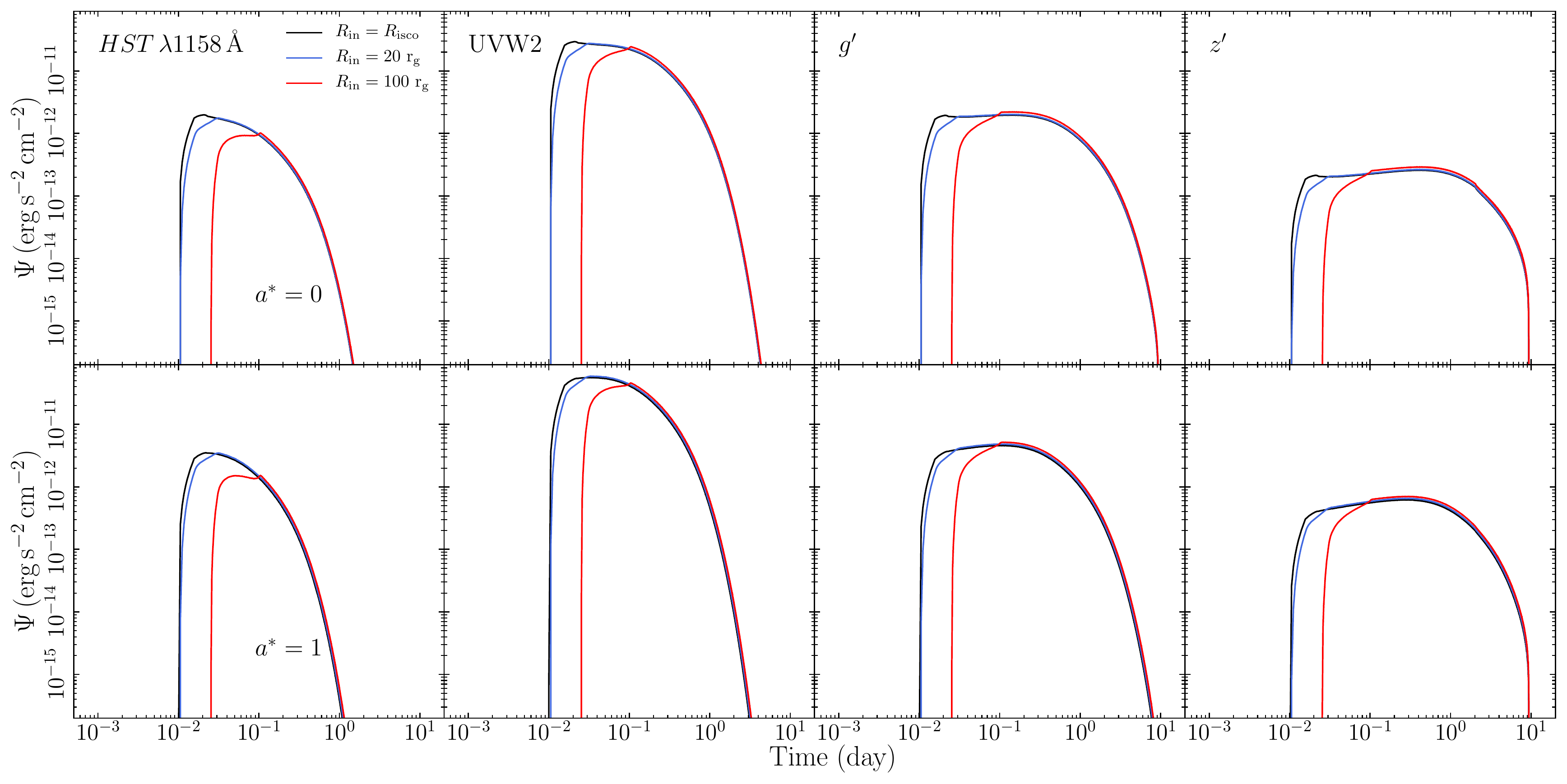}
\caption{Same as Figure \ref{fig:mass}, for the inner radius of the disk.}
\label{fig:Rin}
\end{figure*}

\begin{figure*}
\centering
\includegraphics[width=0.95\linewidth]{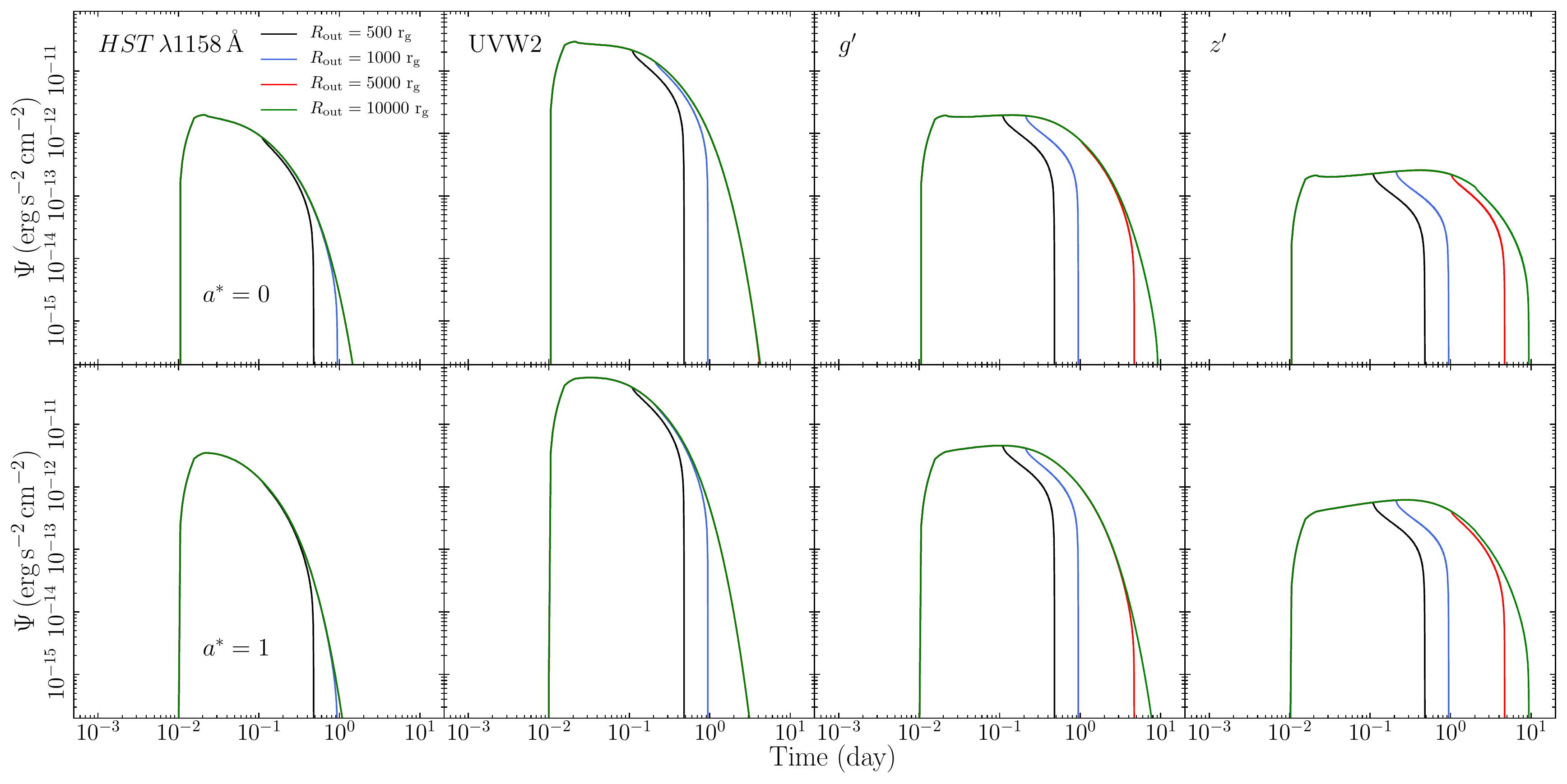}
\caption{Same as Figure \ref{fig:mass}, for the outer radius of the disk.}
\label{fig:Rout}
\end{figure*}

\begin{figure*}
\centering
\includegraphics[width=0.95\linewidth]{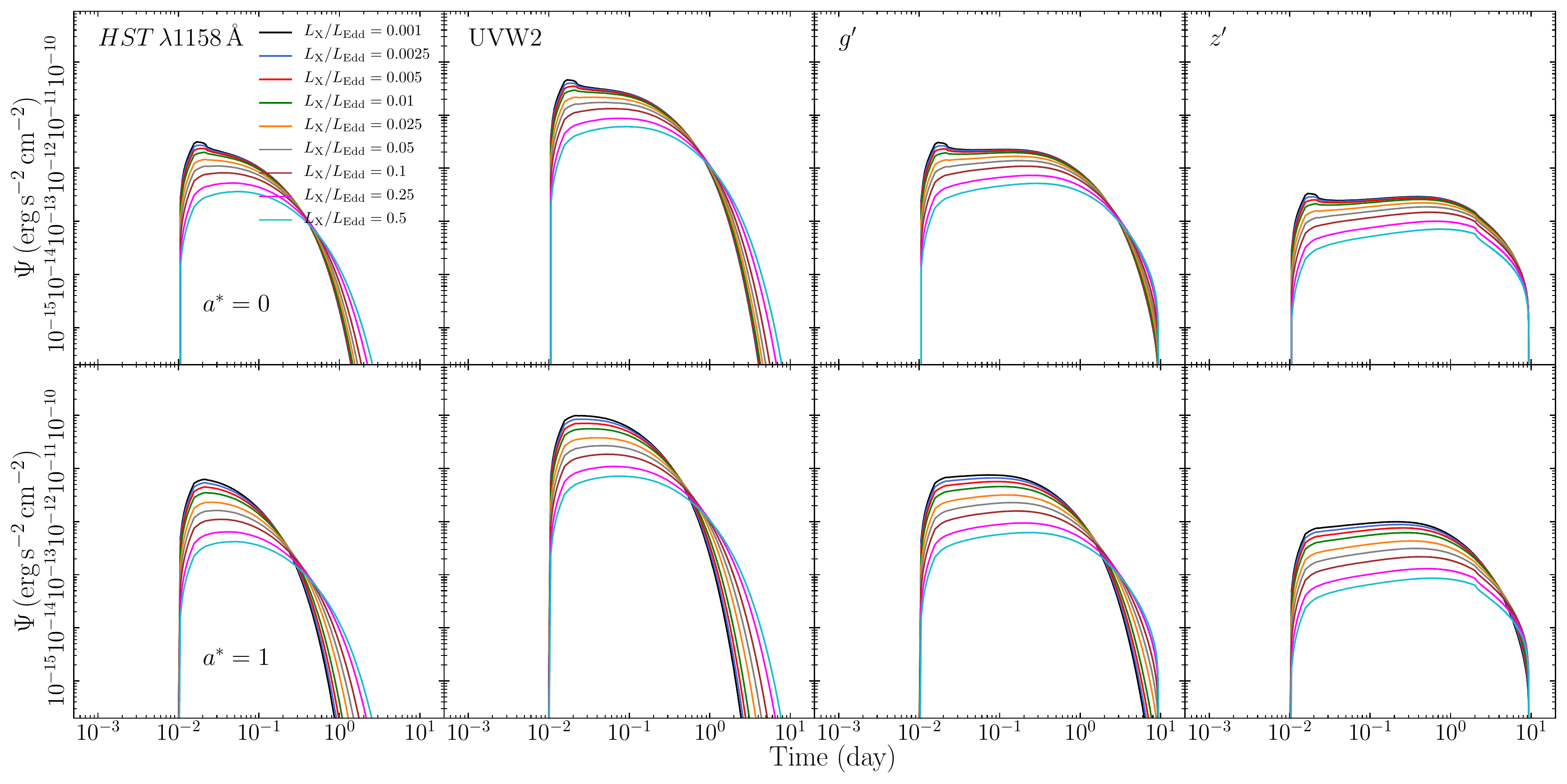}

\caption{Same as Figure \ref{fig:mass}, for the X-ray luminosity.}
\label{fig:Lx}
\end{figure*}

\begin{figure*}
\centering
\includegraphics[width=0.95\linewidth]{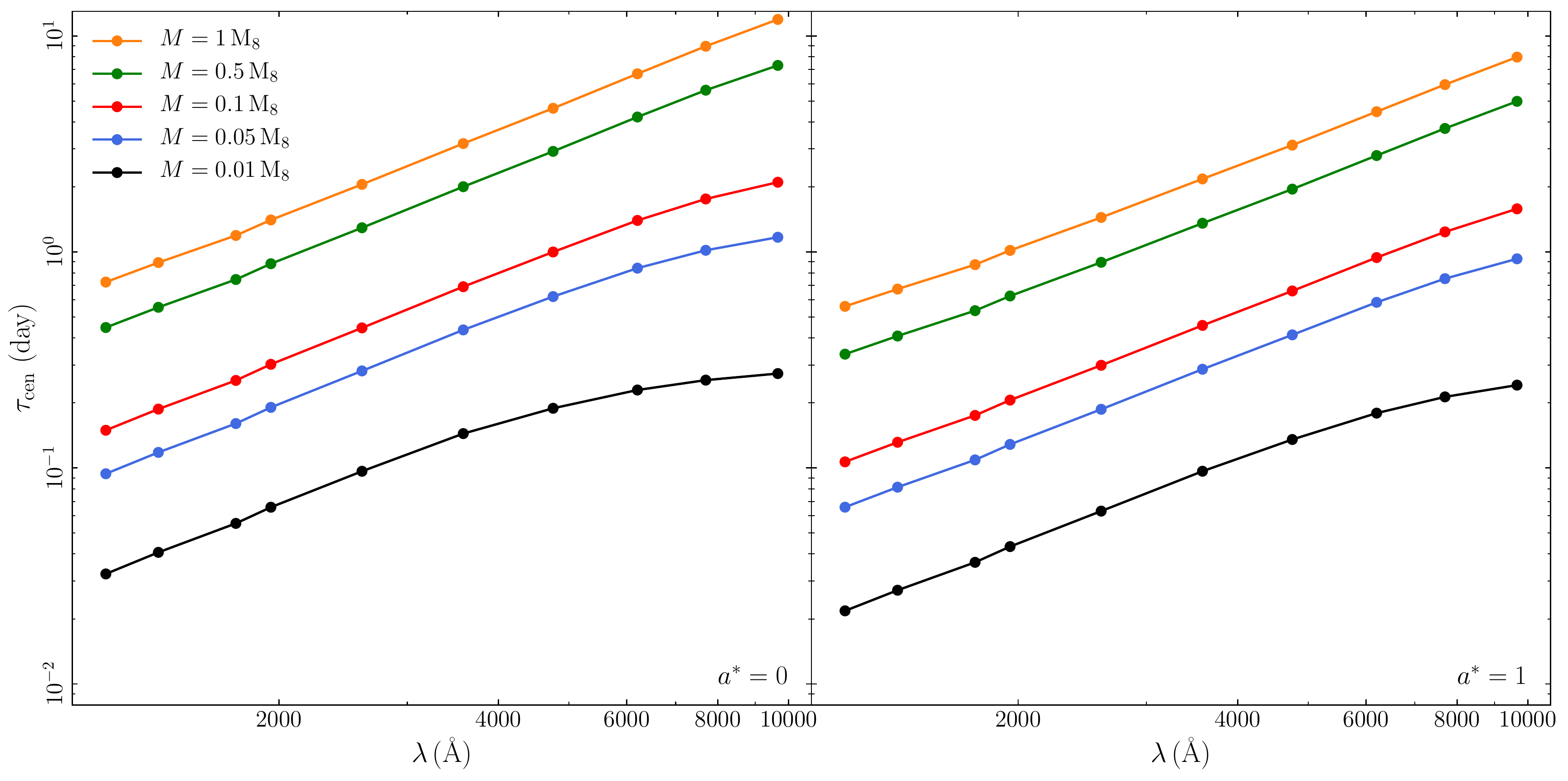}
\caption{The centroid of the CCF (\tcen) 
as a function of wavelength, for the different values of 
the BH mass, considering spins of 0 and 1 (left and right panels, respectively).} 
\label{fig:tau_mass}
\end{figure*}

\begin{figure}
\centering

\includegraphics[width=0.95\linewidth]{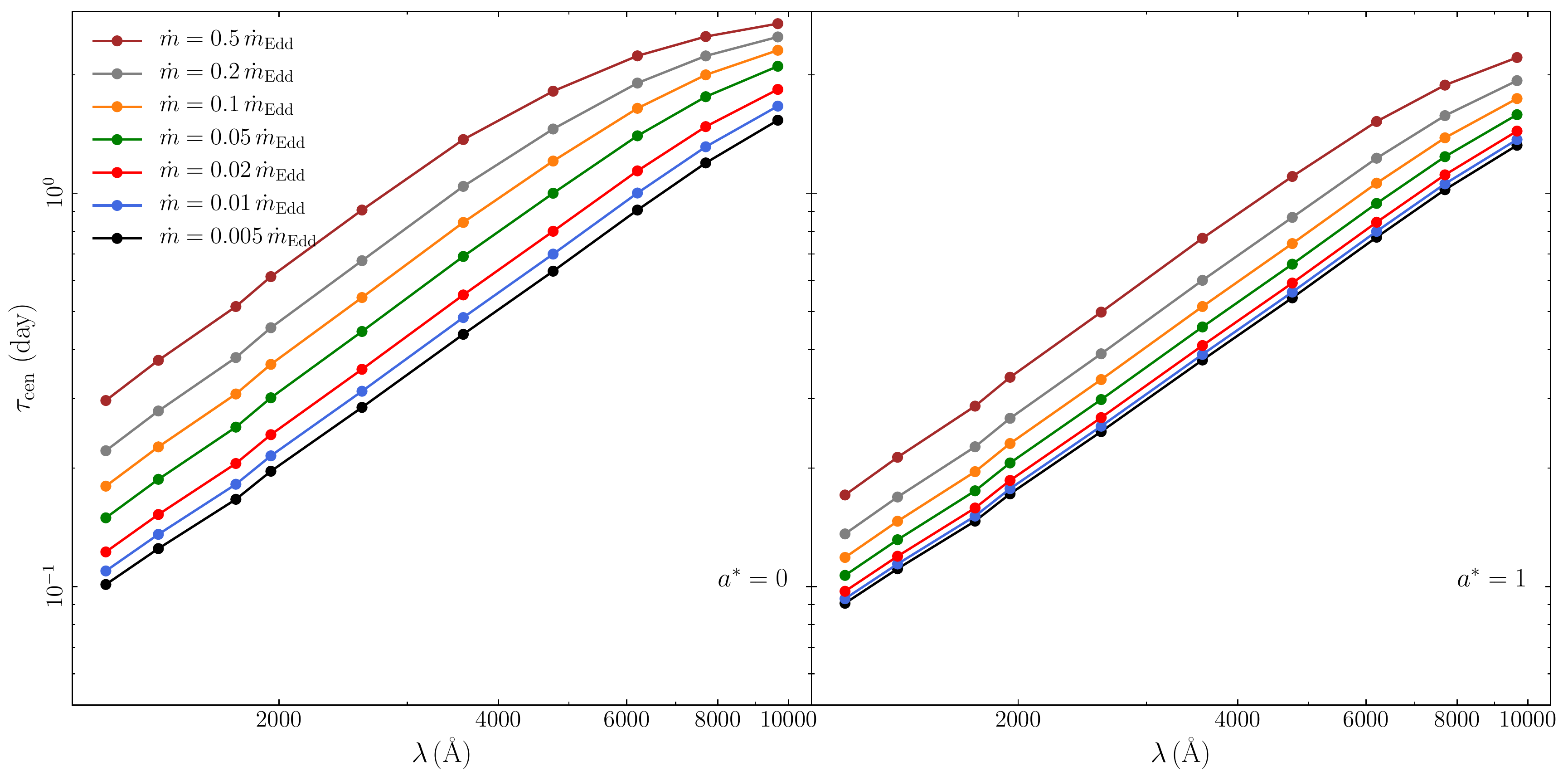}
\caption{Same as Figure \ref{fig:tau_mass} but for the accretion rate.}
\label{fig:tau_mdot}
\end{figure}

\begin{figure*}
\centering
\includegraphics[width=0.95\linewidth]{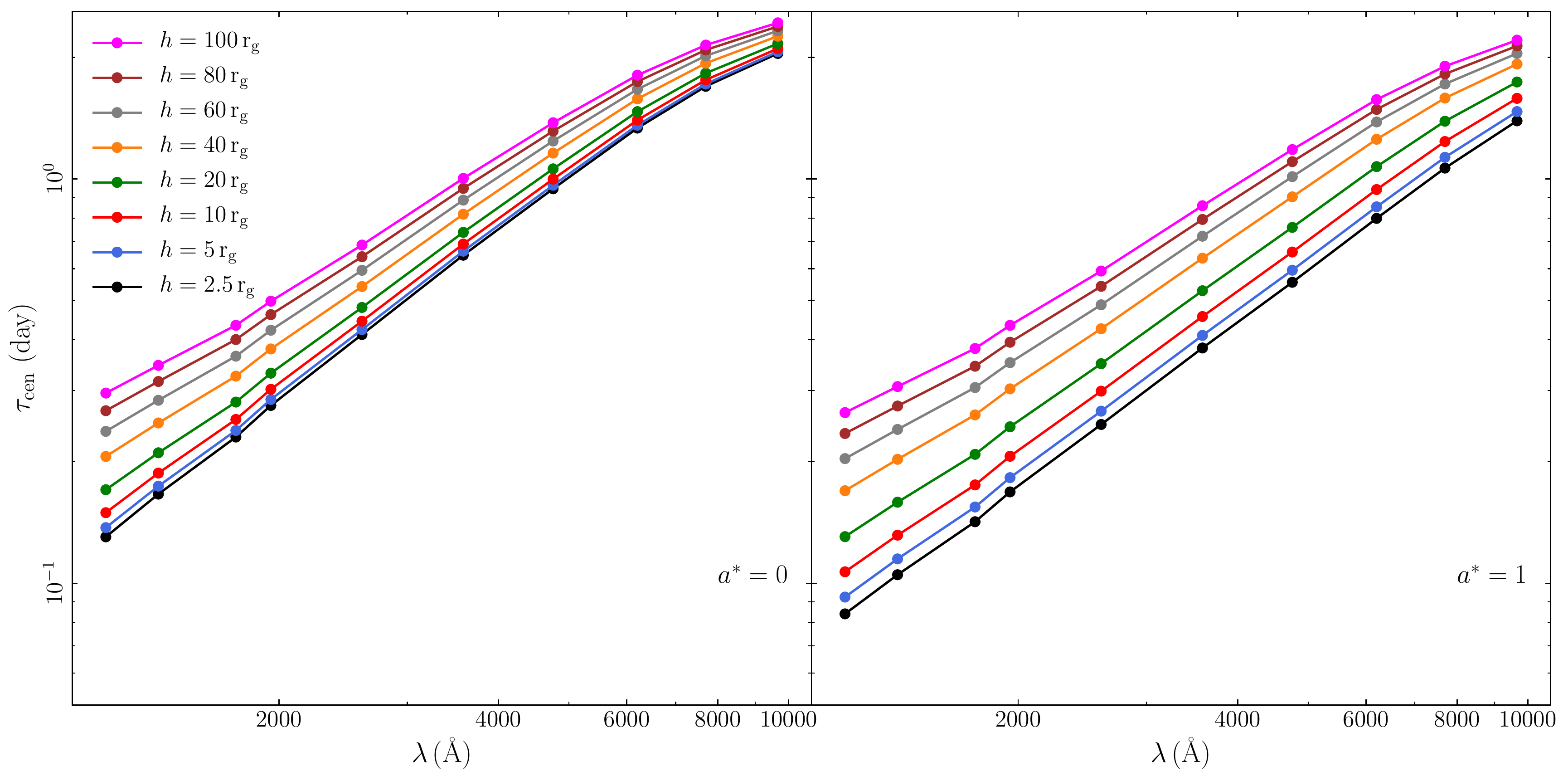}
\caption{Same as Figure \ref{fig:tau_mass} but for the lamp-post height.}
\label{fig:tau_height}
\end{figure*}

\begin{figure*}
\centering

\includegraphics[width=0.95\linewidth]{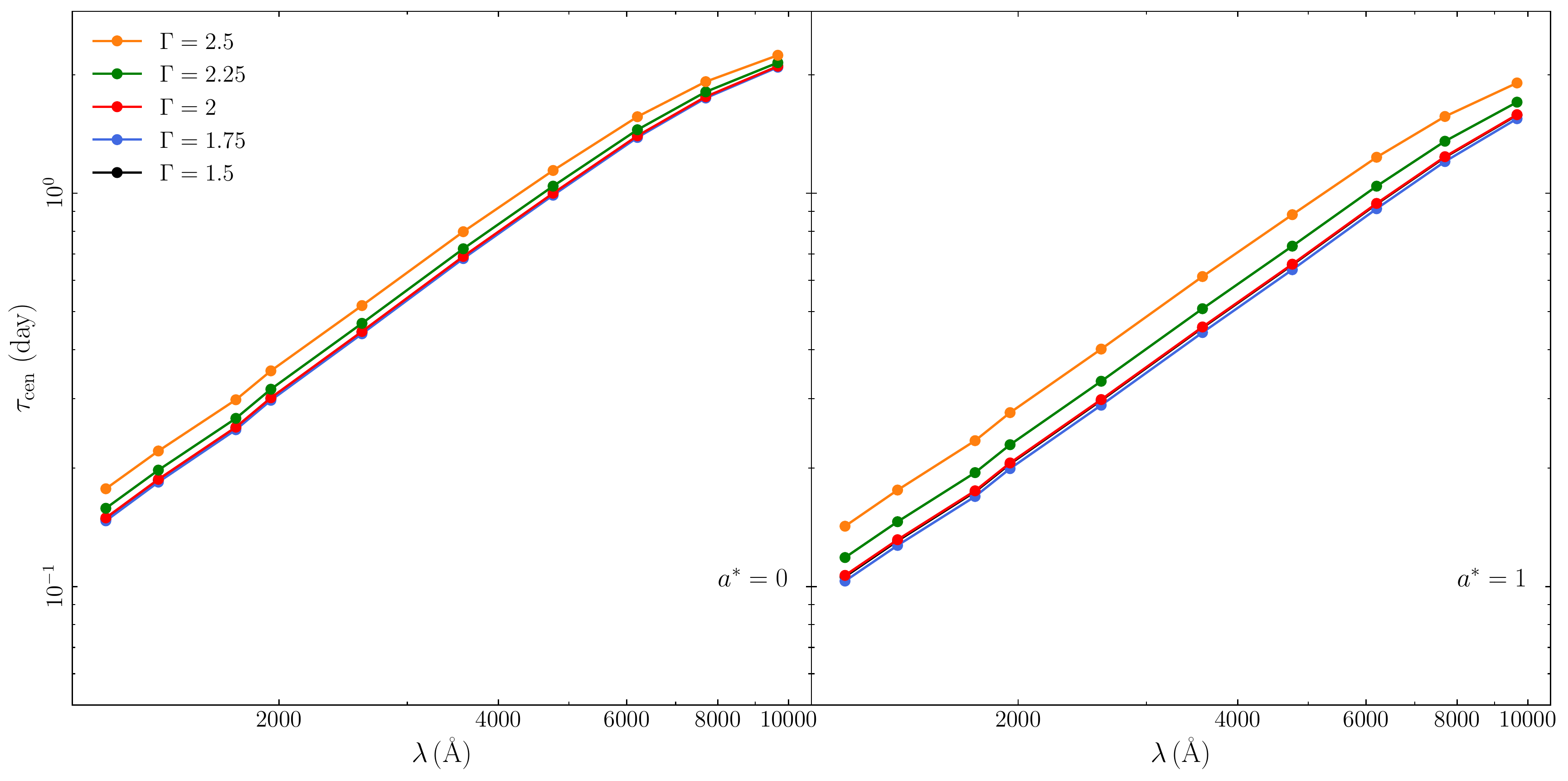}
\caption{Same as Figure \ref{fig:tau_mass} but for the photon index.}
\label{fig:tau_gamma}
\end{figure*}

\begin{figure*}
\centering

\includegraphics[width=0.95\linewidth]{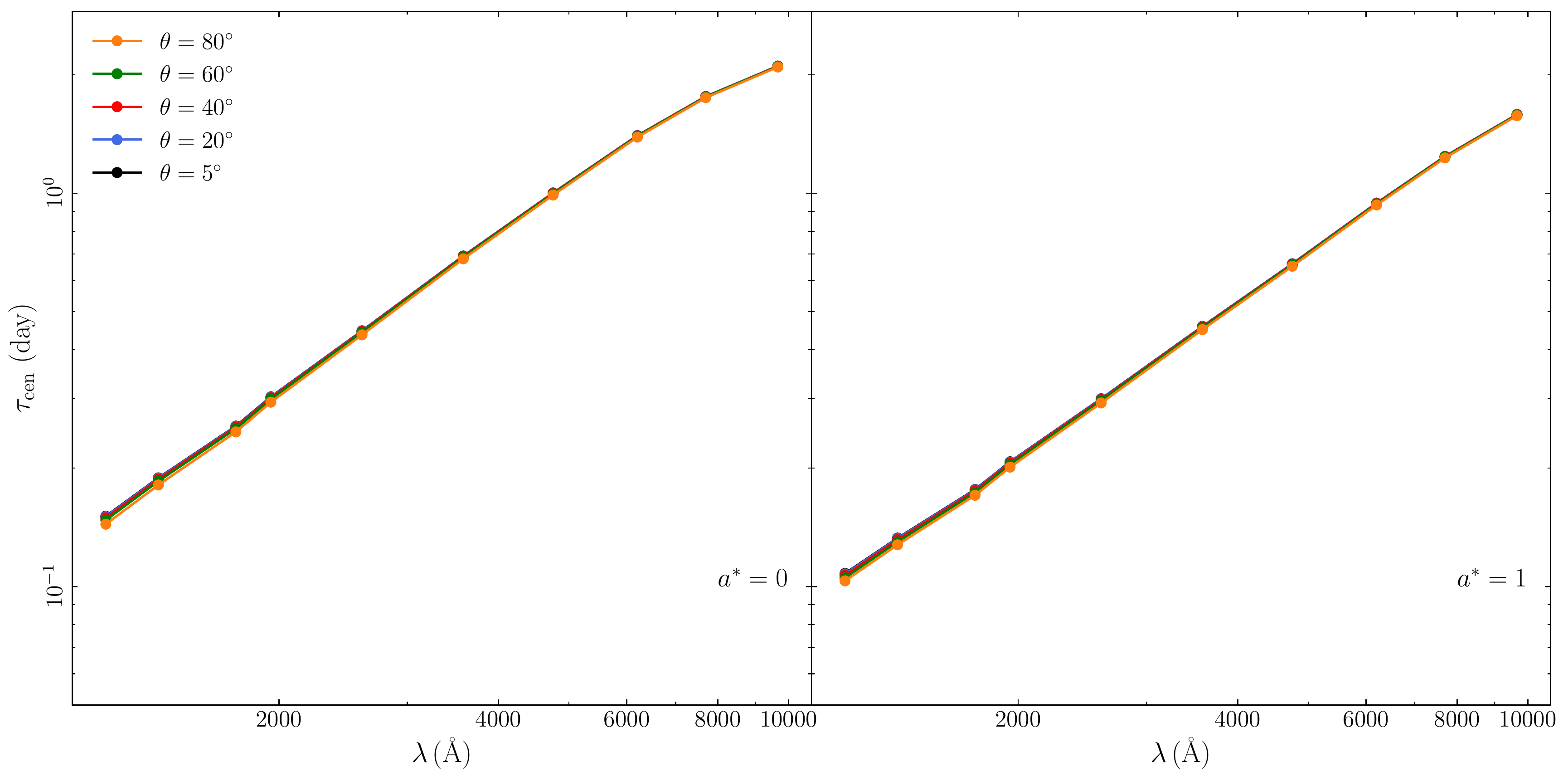}
\caption{Same as Figure \ref{fig:tau_mass} but for the inclination angle.}
\label{fig:tau_theta}
\end{figure*}

\begin{figure*}
\centering

\includegraphics[width=0.95\linewidth]{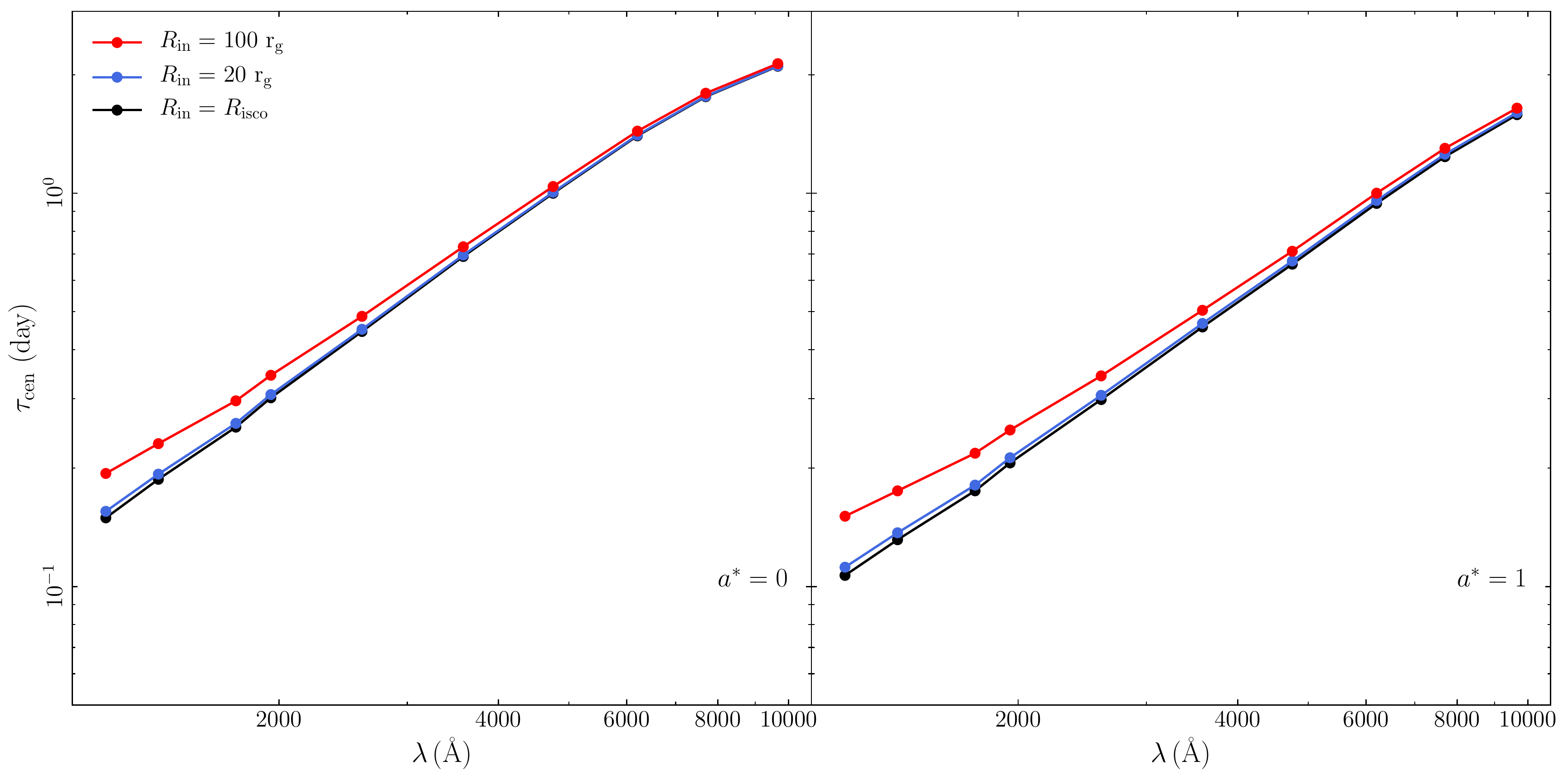}
\caption{Same as Figure \ref{fig:tau_mass} but for the inner radius of the disk.}
\label{fig:tau_Rin}
\end{figure*}

\begin{figure*}
\centering

\includegraphics[width=0.95\linewidth]{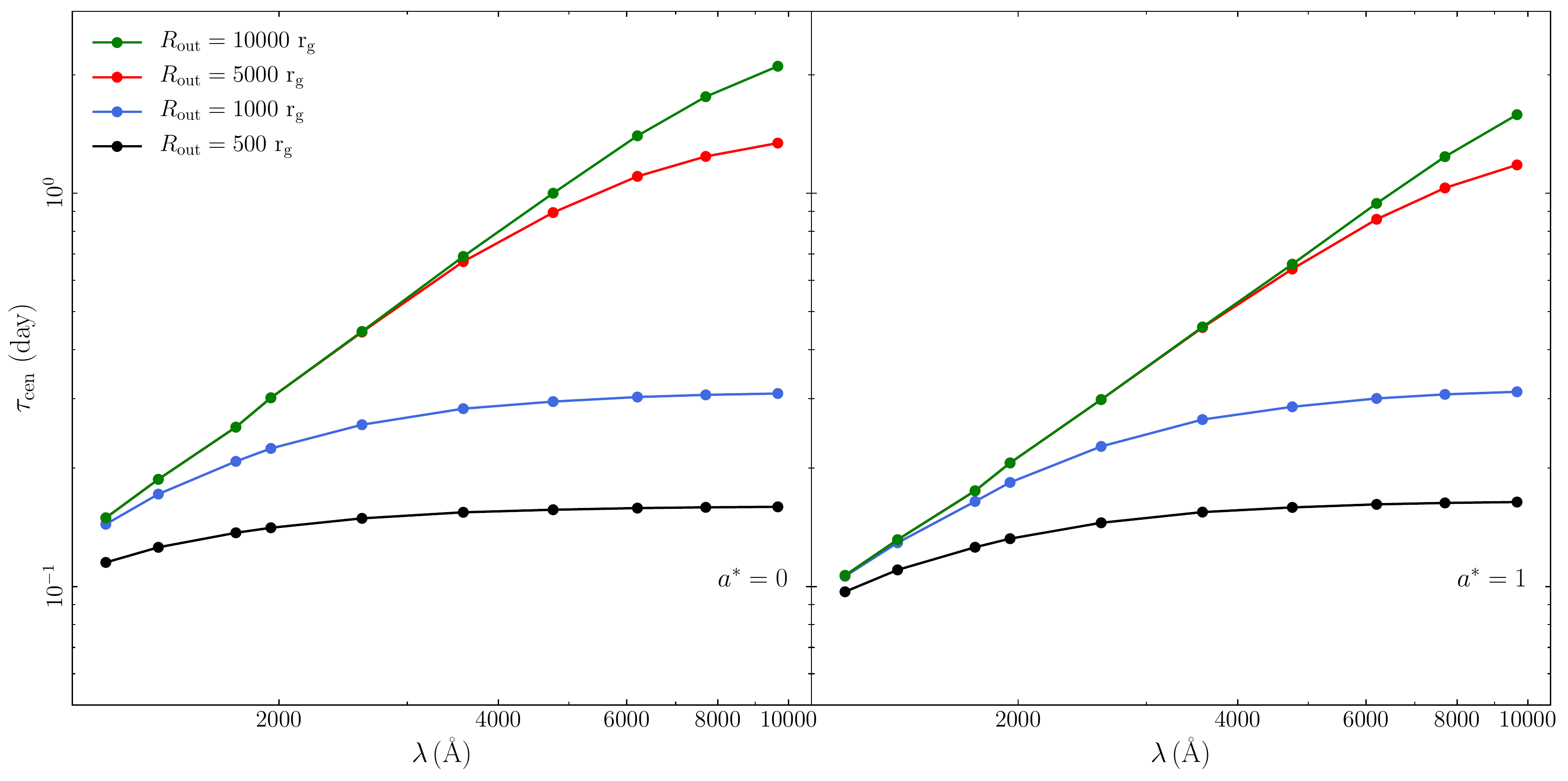}
\caption{Same as Figure \ref{fig:tau_mass} but for the outer radius of the disk.}
\label{fig:tau_Rout}
\end{figure*}

\begin{figure*}
\centering

\includegraphics[width=0.95\linewidth]{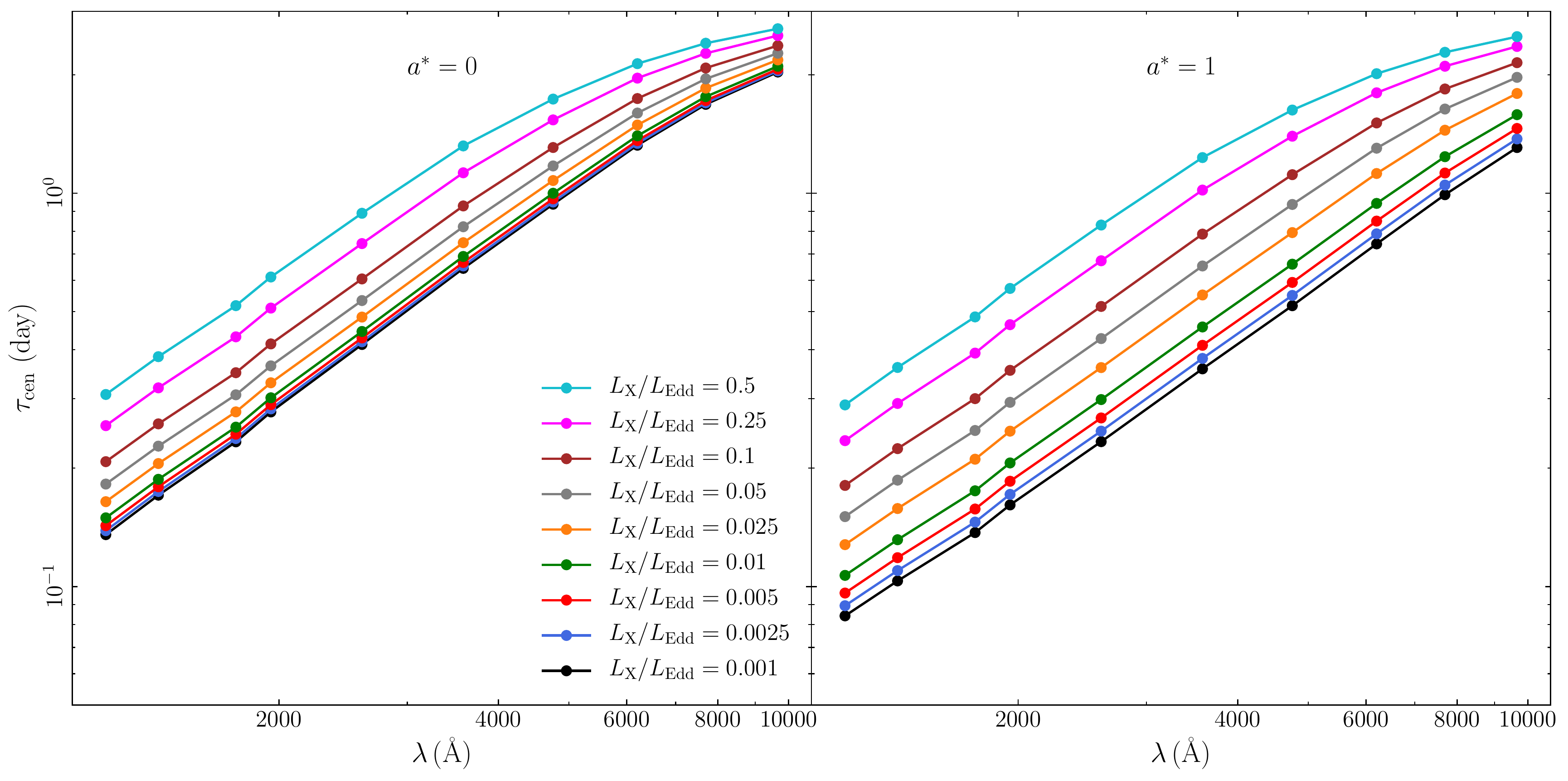}
\caption{Same as Figure \ref{fig:tau_mass} but for the X-ray luminosity.}
\label{fig:tau_Lx}
\end{figure*}

\begin{figure*}
\centering
\includegraphics[width=0.95\linewidth]{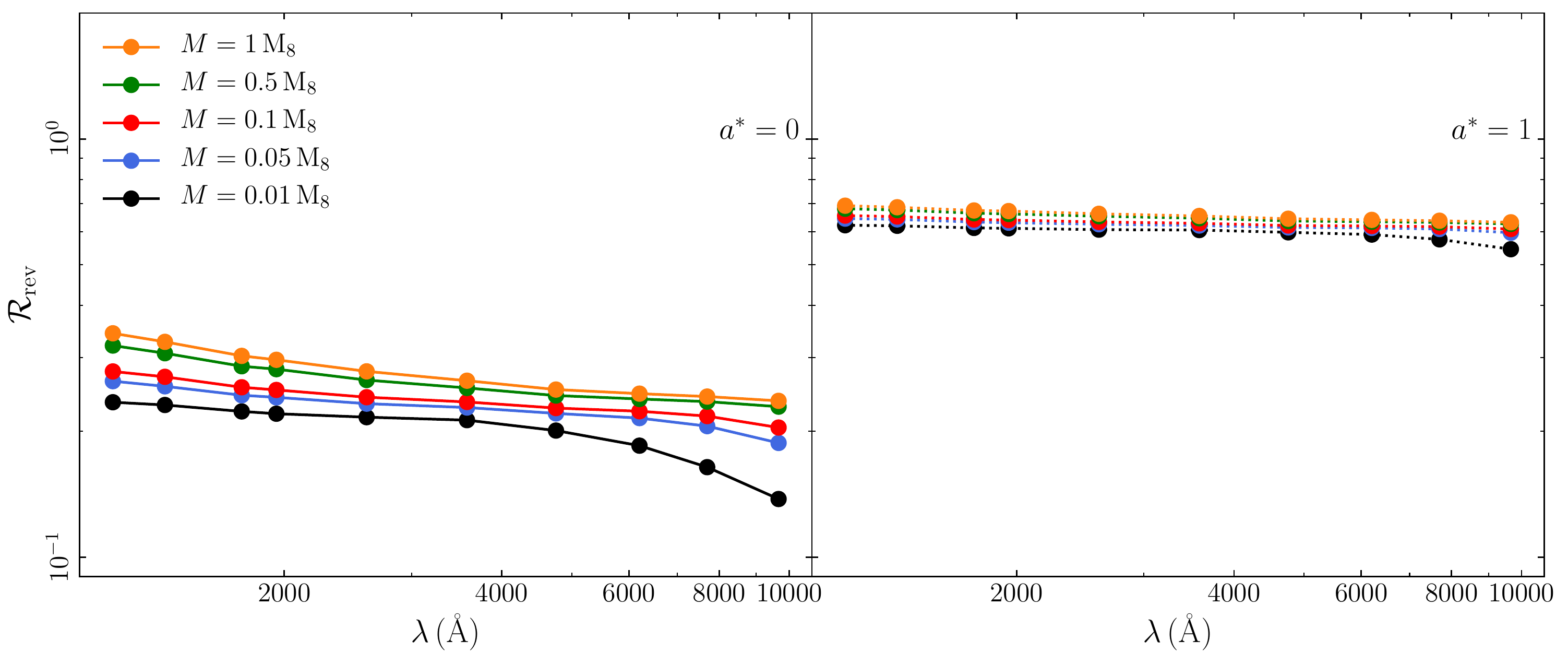}
\caption{The reverberation fraction plotted as a function of the wavelength, for the different values of BH mass, considering spins of 0 and 1 (left and right panels, respectively).}
\label{fig:frev_mass}
\end{figure*}

\begin{figure}
\centering
\includegraphics[width=0.95\linewidth]{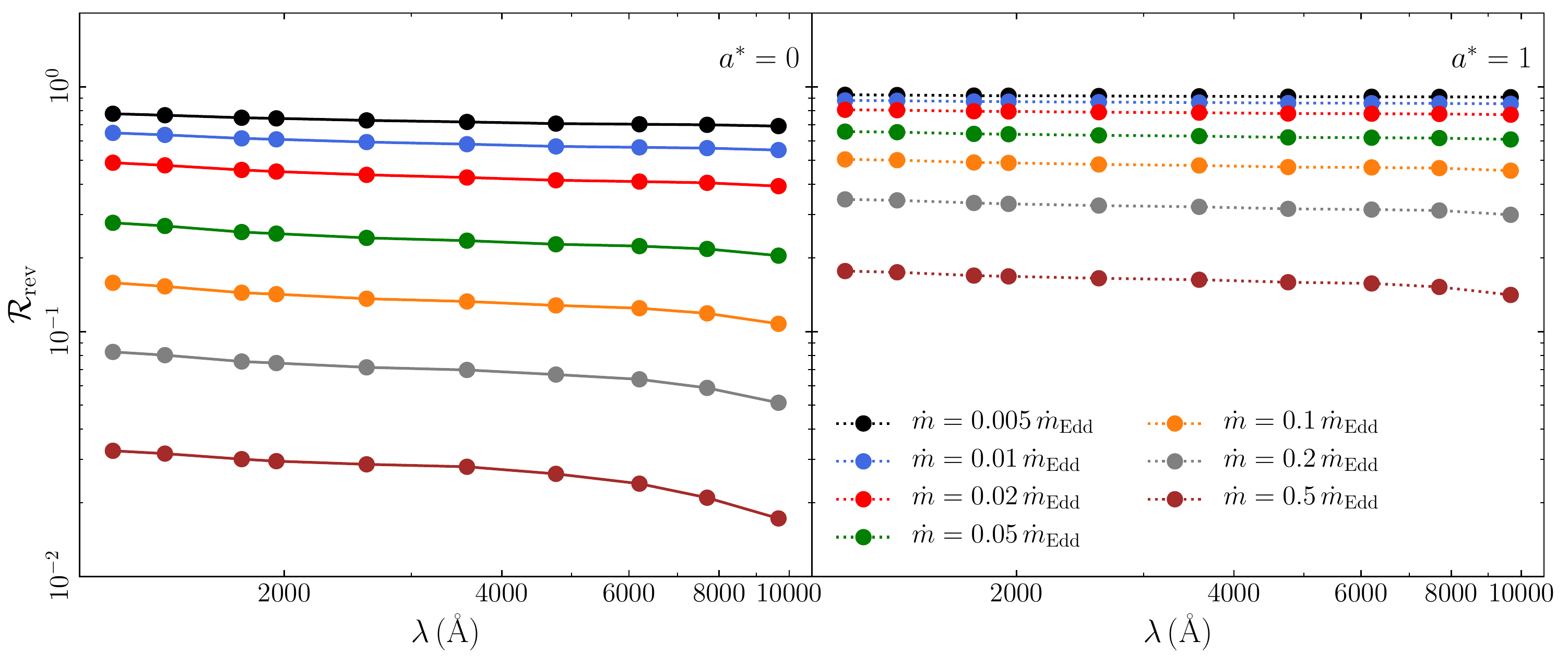}
\caption{Same as Figure \ref{fig:frev_mass} but for the accretion rate.}
\label{fig:frev_mdot}
\end{figure}

\begin{figure*}
\centering
\includegraphics[width=0.95\linewidth]{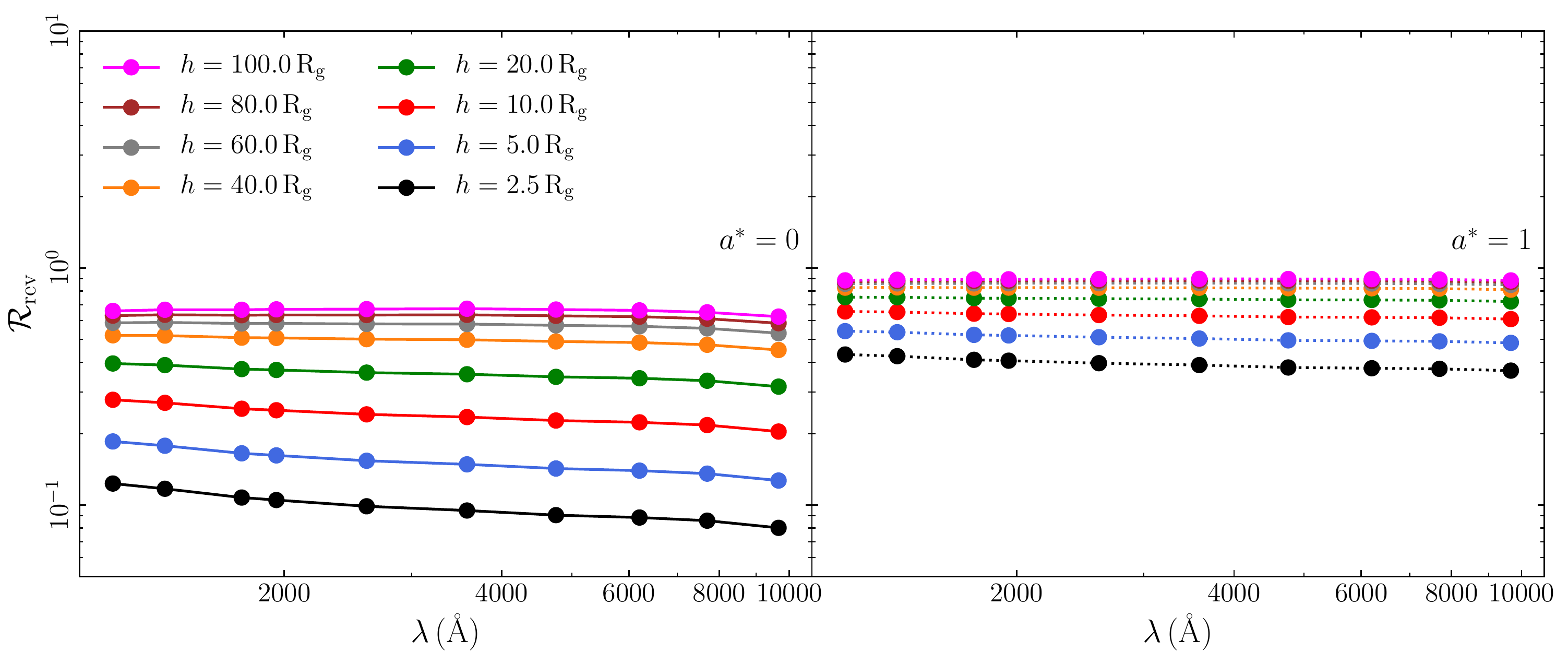}
\caption{Same as Figure \ref{fig:frev_mass} but for the lamp-post height.}
\label{fig:frev_height}
\end{figure*}

\begin{figure*}
\centering
\includegraphics[width=0.95\linewidth]{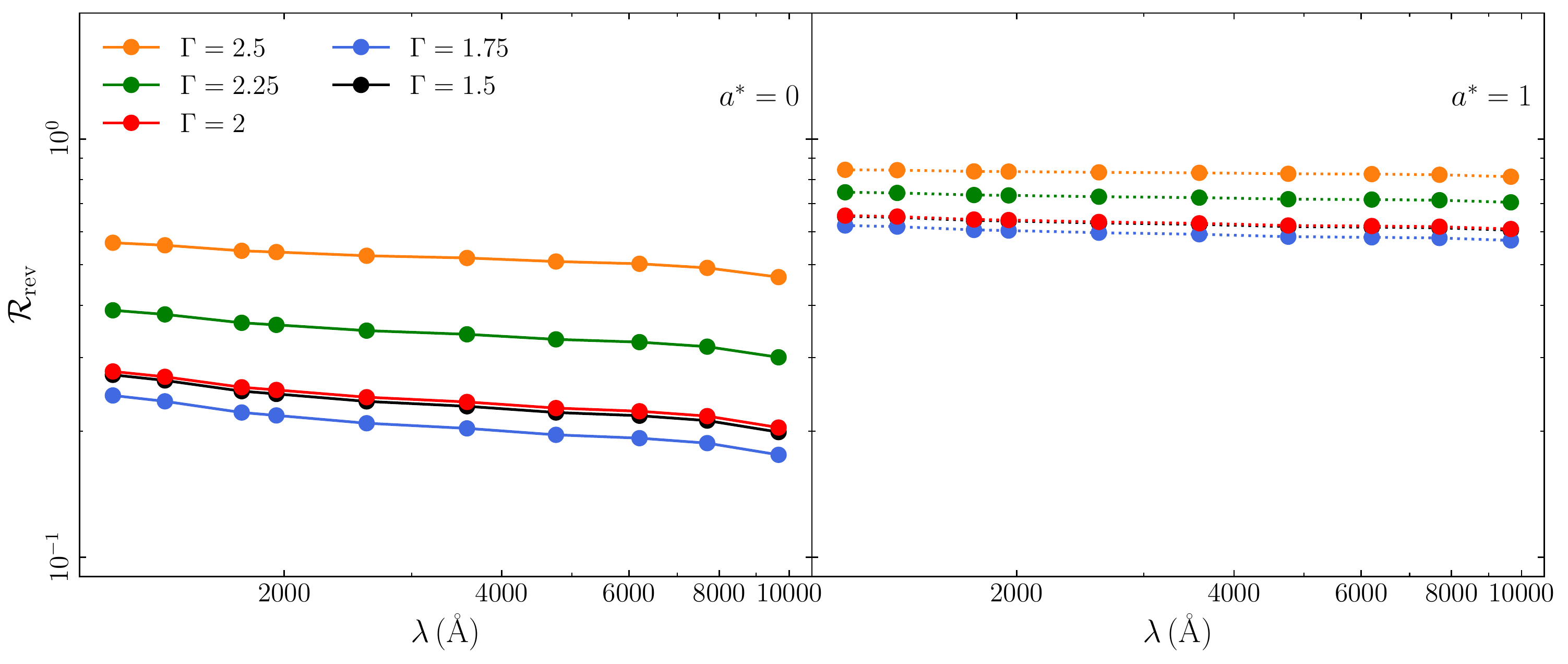}
\caption{Same as Figure \ref{fig:frev_mass} but for the photon index.}
\label{fig:frev_gamma}
\end{figure*}

\begin{figure*}
\centering
\includegraphics[width=0.95\linewidth]{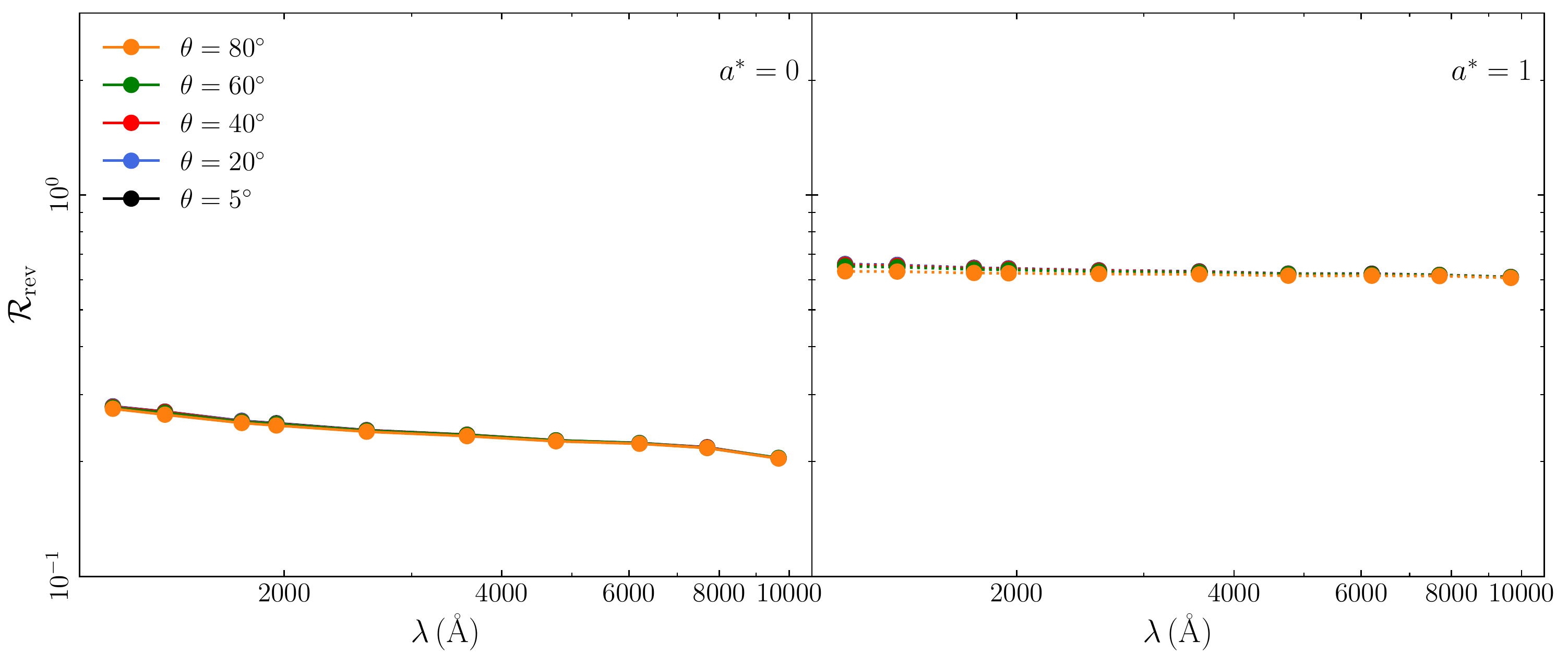}
\caption{Same as Figure \ref{fig:frev_mass} but for the inclination angle.}
\label{fig:frev_theta}
\end{figure*}

\begin{figure*}
\centering
\includegraphics[width=0.95\linewidth]{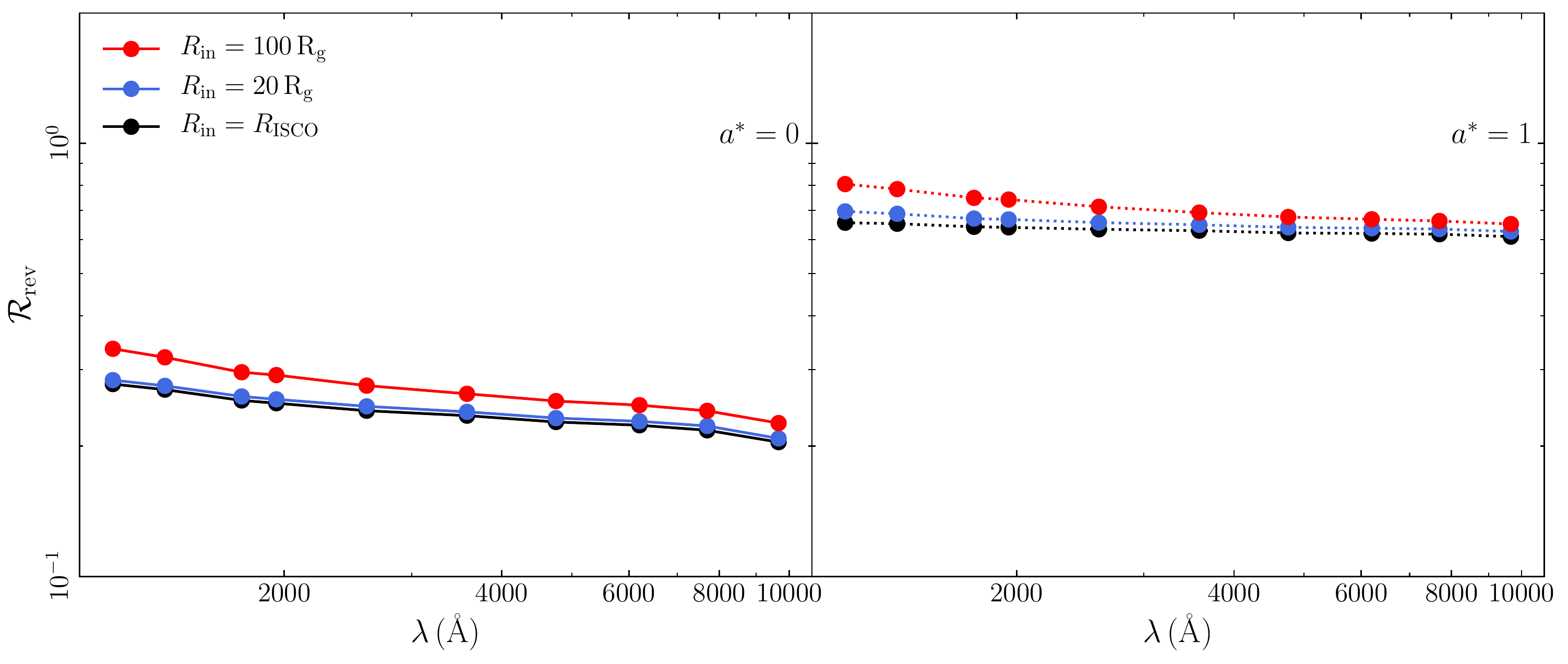}

\caption{Same as Figure \ref{fig:frev_mass} but for the inner radius of the disk.}
\label{fig:frev_Rin}
\end{figure*}

\begin{figure*}
\centering
\includegraphics[width=0.95\linewidth]{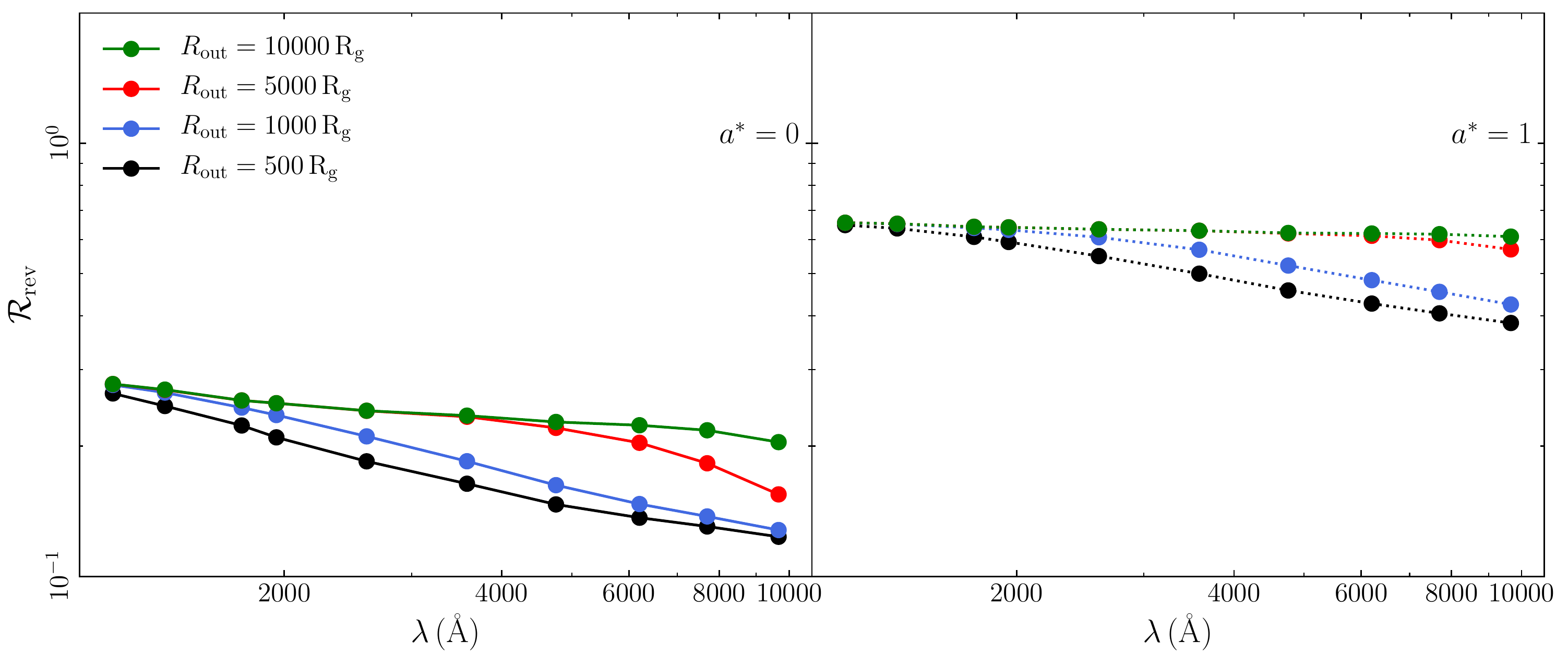}
\caption{Same as Figure \ref{fig:frev_mass} but for the outer radius of the disk.}
\label{fig:frev_Rout}
\end{figure*}

\begin{figure*}
\centering
\includegraphics[width=0.95\linewidth]{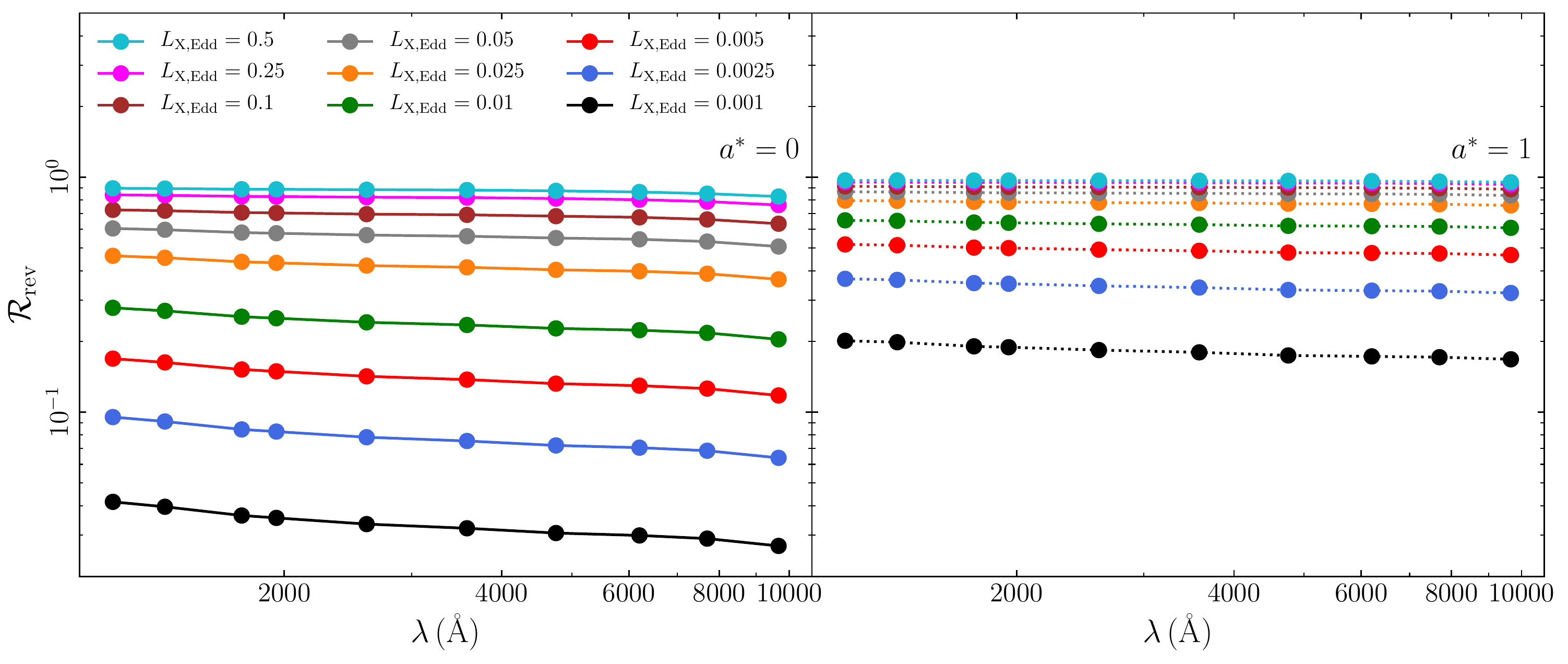}
\caption{Same as Figure \ref{fig:frev_mass} but for the X-ray luminosity}
\label{fig:frev_Lx}
\end{figure*}

In this Section we present the plots of the response functions  (Figures~\ref{fig:mass}-\ref{fig:Lx}), the time-lag spectra  (Figures~\ref{fig:tau_mass}-\ref{fig:tau_Lx}), and the reverberation fraction (Figures~\ref{fig:frev_mass}-\ref{fig:frev_Lx}) for different parameters.

\end{document}